\documentclass[aps,prd,twocolumn,groupedaddress]{revtex4-1}
\usepackage{graphicx}
\include{amssym}

\begin{document}
\title{Proper-time evaluation of the effective action: unequal masses in the loop}
\author{A.\,A.\,Osipov\footnote{Email address: osipov@nu.jinr.ru}}
\affiliation{Joint Institute for Nuclear Research, Bogoliubov Laboratory of Theoretical Physics, 141980 Dubna, Russia}

\begin{abstract}
The Fock-Schwinger proper-time method is used to derive the effective action in the field theory with the chiral $U(3)\times U(3)$ symmetry explicitly broken by unequal masses of heavy particles. The one-loop effective action is presented as a series in inverse powers of heavy masses. The first two Seeley-DeWitt coefficients of this expansion are explicitly calculated. This powerful technique opens a promising avenue for studying explicit flavor symmetry breaking effects in the effective field theories.
\end{abstract}

\maketitle

\section{Introduction}
The heat kernel technique \cite{Hadamard:32} has been introduced to quantum theory in works of Fock \cite{Fock:37a,Fock:37b} and later has been pushed forward by Nambu \cite{Nambu:50} and Schwinger \cite{Schwinger:51}. In combination with the background field method, this allowed DeWitt to develop the manifestly covariant approach to gauge field theory \cite{DeWitt:65}, and quantum gravity \cite{DeWitt:75}. The method allows the nonlocal extension \cite{Barvinsky:87,Barvinsky:90}. It has been widely used in QCD to construct effective meson Lagrangians \cite{Wadia:85,Ebert:86}, in chiral gauge theories to study chiral anomalies \cite{Ball:89}, in cosmology to calculate geometric entropy \cite{Callan:94}, in QED to find Casimir energies and forces \cite{Bordag:01}, etc. 

In all above mentioned cases and in many others \cite{Vassilevich:03}, it becomes necessary to calculate the determinant of a positive definite elliptic operator that describes quadratic fluctuations of quantum fields in the presence of some external or background fields and which in a compact form contains all the information about the one-loop contribution of quantum fields. The result is an asymptotic expansion for the effective action of the theory in powers of the proper-time with Seeley-DeWitt coefficients $a_n(x,y)$. These coefficients are polynomials in the background fields and describe, in the coincidence limit $y\to x$, the local vertices of the corresponding effective Lagrangian. It is remarkable that each term of the expansion is invariant under the action of the group of the internal symmetry, if the theory possesses this symmetry. This follows from the general covariance of the formalism. 

When quantum fields have large equal masses $m$, it is easy to resort to an expansion in the inverse powers of mass, which is valid when all background fields and their derivatives are small compared to the mass of the quantum field. In this case, the asymptotic coefficients $a_n$ do not change. Such long wavelengths ($\lambda\gg 1/m$) expansion allows one to obtain an action that takes into account effectively the leading low energy effect of virtual heavy states. This scenario is realized in theories with spontaneously broken symmetry, or in the theories with a large bare mass. A typical example of the first type is the Nambu -- Jona-Lasinio (NJL) model \cite{Nambu:61a,Nambu:61b}, where ground state in the strong coupling regime is found to be separated by a gap from the excited states (quasiparticles), which are identified with nucleons. Reinterpreting nucleons as quarks, one obtains a low energy meson action from one-loop quark dynamics \cite{Volkov:84}. The proper-time method is especially useful here \cite{Wadia:85,Ebert:86}. Examples of the second type arise under extension of some effective field theory $X$ with symmetry group $G$ to the other effective theory $X'$, when $X'$ contains heavy degrees of freedom belonging to some representation of $G$. At present, such theories are being actively studied in the context of extending the standard model of electroweak interactions \cite{Passarino:19}.  

In realistic models one is often confronted with the difficulty that the flavor symmetry is broken by large unequal masses $M=\mbox{diag} (m_1, m_2, \ldots ,m_f)$. In this case the complete factorization of $M$ in the heat kernel is impossible because $M$ does not commute with the rest of an elliptic operator. The consequence is that the Seeley-DeWitt coefficients depend in a complex way on both the fields and the mass-dependent-constants of their interactions. The treatment of such case is known to be an onerous task, especially when both Bose and gauge fields are present in the elliptic operator \cite{Min:82,Min:89a,Min:89b}.

Recently \cite{Osipov:21a,Osipov:21b}, a new algorithm based on the proper-time method has been proposed for deriving the effective action in a theory with heavy virtual fermions (or bosons) of unequal masses belonging to some representation of the symmetry group $G$. These short Letters contain the main idea and the final result. Non-trivial calculation details have been omitted due to their complexity. The purpose of the present contribution is to supply all necessary details of such non-trivial calculations, without knowing which it is difficult to be convinced in the validity of the previously stated results. 

The new algorithm generalizes the standard large mass expansion of the heat kernel to the case of unequal masses by the formula 
\begin{equation}
\label{alg}
e^{-t(M^2+A)}=e^{-tM^2}\!\left[1+\sum_{n=1}^\infty (-1)^n f_n(t,A) \right],
\end{equation}
where $M=\mbox{diag} (m_1, m_2, \ldots ,m_f)$ is a diagonal mass matrix; $t$ is the proper-time parameter; the expression in the square brackets is the time-ordered exponential $\mbox{OE}[-A](t)$ of $A(s)= e^{sM^2}\! A\, e^{-sM^2}$, and $A$ is a positive definite self-adjoint elliptic operator in some background (its explicit form will be clarified later), accordingly 
\begin{equation}
f_n(t,A)=\!\int\limits_0^t\!\! ds_1\!\!\int\limits_0^{s_1}\!\! ds_2 \ldots \!\!\!\!\int\limits_0^{s_{n-1}}\!\!\!\! ds_n A(s_1) A(s_2) \ldots A(s_n).
\end{equation} 
If masses are equal, this formula yields the well-known large mass expansion with standard Seeley-DeWitt coefficients $a_n(x,y)$ \cite{Ball:89}. In fact, formula (\ref{alg}) is an extension of the Schwinger's method used to isolate the divergent aspects of a calculation in integrals with respect to the proper-time \cite{Schwinger:51,DeWitt:75} to the non-commutative algebra. 

There is a simple heuristic argument that explains why this formula is also relevant for describing the generalized $1/M$ series. Indeed, the $1/M$ expansion is known to be valid when all background fields and their derivatives are small compared to the mass of quantum fields. Therefore, factoring $e^{-tM^2}$ one separates the leading contribution. The remaining part of the heat kernel may be unambiguously evaluated by expanding it in a power series in $t$ about $t=0$. As a consequence, the Seeley-DeWitt coefficients $a_n$ receive corrections: $a_n\to b_n=a_n+\Delta a_n$, where $\Delta a_n$ vanish in the limit of equal masses. 

Currently, there are two methods for deriving quantum corrections induced by virtual states of unequal masses. In \cite{Min:82,Min:89a,Min:89b}, the heat kernel is evaluated on the bases of the modified DeWitt WKB form. 
This yields a different asymptotic series for the right-hand side of Eq.\,(\ref{alg}), and, consequently, the different expressions for $\Delta a_n$. The approach proposed in \cite{Osipov:01a,Osipov:01b,Osipov:01c} starts from the formula (\ref{alg}), but afterwards an additional resummation of the asymptotic series is applied. This essentially simplifies the calculations, but changes the structure of the $1/M$ series. As a result, one looses correspondence between a mass-dependent factor at the effective vertex and a flavor content of the one-loop Feynman diagram which generates the vertex. Here I abandon this resummation.
 
The utility of the proper-time technique is that it reduces the task of the large mass expansion to a simple algebraic problem which requires less work than one needs for the corresponding Feynman diagrams calculation in momentum space. In the following, we consider a quite non-trivial case of the chiral $U(3)\times U(3)$ symmetry broken by the diagonal mass matrix $M=\mbox{diag} (m_1, m_2, m_3)$ to demonstrate the power of the method. To find the two leading contributions $b_1(x,x)$ and $b_2(x,x)$ in the $1/M$ expansion of the effective action one requires to consider only four terms of the series (\ref{alg}) that results in more than a hundred effective vertices. 

The effects of flavor symmetry breaking are currently important in many physical applications: in studies of physics beyond standard model to construct the low energy effective action by integrating out the heavy degrees of freedom \cite{Passarino:19,Bizot:18}; in two Higgs doublet models \cite{Branco:12} to address the problem of almost degenerate Higgs states at $125\, \mbox{GeV}$ \cite{Bian:18,Haber:19}; in the low energy QCD to study the $SU(3)$ and isospin symmetry breaking effects \cite{Taron:97}. These effects are known to be very important to explore the QCD phase diagram \cite{Osipov:13}, to study a formation of the strange-quark matter \cite{Witten:84,Osipov:15}, to study nuclear matter in extreme conditions that arose in nature at the early stages of the evolution of the Universe and in the depths of neutron stars \cite{Chernodub:11,Gatto:11}. The method described below, in particular, can be used for obtaining low energy meson effective Lagrangian in the framework of the NJL model, as an alternative to the approaches developed in \cite{Ebert:86,Volkov:86}.

The paper is organized as follows. In Sec.\,\ref{P-t-exp}, we formulate the method and present the basic steps required to construct the desired $1/M$ expansion. The Sec.\,\ref{L-t-exp} sets out the necessary details related to the calculation of the first two coefficients of the asymptotic series. In Sec.\,\ref{C-M-s}, we present the effective Lagrangian describing the self-energy and interactions of zero-spin and one-spin bosons in Minkowski space-time induced by the intermediate one-loop quark diagrams. A short summary and some concluding remarks are given in Sec.\,\ref{Conclusions}. Many important technical points related to our calculations are collected in six Appendices.

\section{Proper-time expansion}
\label{P-t-exp}
In this section we obtain the $1/M$ series of the effective one-loop action in Euclidian space and explicitly single out the structures necessary for calculating the first two coefficients of this asymptotic expansion.

\subsection{Determinant of the Dirac operator}
The logarithm of the formal determinant of the self-adjoint elliptic operator of the second order describes the one-loop radiative corrections to classical theory. In the following, we are interested in the real part of the effective action resulting from the calculation of the determinant of the Dirac operator $D$ in the background scalar $s$, pseudoscalar $p$, vector $v_\mu$, and axial-vector $a_\mu$, fields. The proper-time method cannot be applied directly to fermions, since the Dirac operator $D$ is a first order elliptic operator, and its spectrum is unbounded both above and below. Instead, one should consider the functional 
\begin{equation}
\label{logdet}  
  W_E =\ln |\det D_E| 
         = -\int\limits^\infty_0\!\frac{dt}{2t}\,\rho_{t,\Lambda}\,\mbox{Tr}\left(e^{-t D_E^\dagger D_E^{}}\right),
\end{equation}
representing a real part of the one-loop effective action in Euclidian space as the integral over the proper-time $t$. Notice that the Hermitian operator $D_E^\dagger D_E^{}$ is a second order elliptic operator, unbounded above, so we can use the proper-time method to regularize it precisely as for bosons. The integral diverges at the lower limit, therefore, a regulator $\rho_{t,\Lambda}$ is introduced, where $\Lambda$ is an ultraviolet cutoff. Since we will carry out calculations in Euclidean space (the subscript $E$ marks this), and the obtained result will be analytically continued to Minkowski space, we will adhere to certain rules of transition from one space to another, which we collect in Appendix \ref{app1}. 
 
For definiteness, suppose that one is dealing with the effective action arising due to integration over light quark degrees of freedom. In this case, the Dirac operator $D_E$ in Euclidean four-dimensional space has the form  
\begin{equation}
D_E^{}=i\gamma_\alpha d_\alpha-M+ s+i\gamma_{5E} p ,
\end{equation}
where $d_\alpha=\partial_\alpha +i\Gamma_\alpha$, $\Gamma_\alpha =v_\alpha +\gamma_{5E}a_\alpha$, $\alpha =1,2,3,4$. The external fields are embedded in the flavor space through the set of matrices $\lambda_a = (\lambda_0, \lambda_i)$, where $\lambda_0 =\sqrt{2/3}$ and $\lambda_i$ are the eight $SU(3)$ Gell-Mann matrices; for instance,  $s=s_a\lambda_a$, and so on for all fields. The quark masses are given by the diagonal matrix $M=\mbox{diag} (m_1, m_2, m_3)$ in the flavor space. The symbol "Tr" denotes the trace over Dirac $(D)$ $\gamma$-matrices, color $(c)$ $SU(3)$ matrices, and flavor $(f)$ matrices, as well as integration over coordinates of the Euclidean space: $\mbox{Tr}\equiv \mbox{tr}_I \int\! d^4x_E$, where $I=(D,c,f)$. The trace in the color space is trivial: it leads to the overall factor $N_c=3$. The dependence on external fields in $D_E$ after switching to the Hermitian operator
\begin{equation}
D_E^\dagger D_E^{}=M^2 -d^2+Y
\end{equation}
is collected in $Y$ and the covariant derivative $d_\alpha$. In the following we do not need an explicit expression for $Y$, 
nevertheless, for completeness, we include this expression in Appendix \ref{app1}.

If quarks were massless, the theory would have a global $U(3)_L \times U(3)_R $ chiral symmetry. It is known, however, that the ground state of QCD is not invariant under the action of chiral group. As a result, the entire system undergoes a phase transition accompanied by the appearance of a gap in the fermion spectrum. Quarks get their equal nonzero constituent masses. 

Additionally, due to explicit breaking of chiral symmetry, realized through the mass terms of current quarks, one can easily find that the inequality of current masses after spontaneous breaking of chiral symmetry leads to the inequality of constituent quark masses.  

Thus, we arrive at a problem in which one needs to study the properties of a system at large scales, i.e., one needs an   expansion of the effective action in the inverse powers of large unequal masses of quarks
\begin{equation}
\label{mass}
   M=\sum_{i=1}^{3} m_i E_i,\quad (E_i)_{jk}=\delta_{ij}\delta_{ik}, \quad E_iE_j=\delta_{ij}E_j.
\end{equation}
The matrix $E_i$ maps the point $(u,d,s)$ of the three-dimensional flavor space to the point $(u,0,0)$, if $i=1$, to the point $(0,d,0)$, if $i=2$, and to the point $(0,0,s)$, if $i=3$. Thus $E_i$ is an orthogonal projection onto the flavor space which can be expressed through the $\lambda$-matrices: 
\begin{eqnarray}
E_1&\equiv&\lambda_u=\frac{1}{\sqrt 6}\lambda_0 +\frac{1}{2}\lambda_3+\frac{1}{2\sqrt 3}\lambda_8\,, \label{Ei:1} \\
E_2&\equiv&\lambda_d=\frac{1}{\sqrt 6}\lambda_0 -\frac{1}{2}\lambda_3+\frac{1}{2\sqrt 3}\lambda_8\,, \label{Ei:2} \\
E_3&\equiv&\lambda_s=\frac{1}{\sqrt 6}\lambda_0 - \frac{1}{\sqrt 3}\lambda_8\,. \label{Ei:3}
\end{eqnarray}  

Notice also the following useful relations which are important for our calculations. The first formula is a direct consequence of the fact that the quark mass is given by the diagonal matrix   
\begin{equation}
e^{-tM^2}=\sum_{i=1}^{3} e^{-tm_i^2} E_i.
\end{equation}
The second formula reflects the projection property of the $E_i$-matrices
\begin{equation}
\label{projprop}
E_i AE_j =A_{ij}E_{ij},
\end{equation}
where $(E_{ij})_{mn}=\delta_{im} \delta_{jn}$, in particular, $E_{ii}=E_i$. It is true for any flavor matrix $A$ whose entries are given by $A_{ij}$. Note that here and in the following we sum over repeated flavor indices only when a symbol of the sum is explicitly written out. The orthogonal basis $E_{ij}$ has the properties
\begin{eqnarray}
\label{trpropEij}
&&\mbox{tr}_f \left(E_{ij}\right)=\delta_{ij}, \nonumber \\
&&\mbox{tr}_f \left(E_{ij}E_{kl}\right)=\delta_{il}\delta_{jk}, \nonumber \\
&&\mbox{tr}_f \left(E_{ij}E_{kl}E_{mn}\right)=\delta_{in}\delta_{jk}\delta_{lm},  \\
&& \ldots\ldots\ldots\ldots\ldots         \nonumber \\
&&\mbox{tr}_f \left(E_{i_1i_2}\ldots E_{i_{2n-1}i_{2n}}\right)=\delta_{i_1i_{2n}}\delta_{i_2i_3}\ldots \delta_{i_{2n-2}i_{2n-1}}. \nonumber 
\end{eqnarray}
 
To advance in the evaluation of expression (\ref{logdet}), we use the Schwinger technique of a fictitious Hilbert space \cite{Schwinger:51}. Then a matrix element of a quantum mechanical operator can be treated as 
\begin{equation}   
   \mbox{Tr}\left(e^{-tD_E^\dagger D_E^{}}\right)\equiv\int\!\! d^4x_E\,\mbox{tr}_I\,\langle x|e^{-tD_E^\dagger D_E^{}}|x\rangle\,.
\end{equation}
The use of a plane wave with Euclidian 4-momenta $k$, $\langle x|k\rangle $, as a basis greatly simplifies the calculations (details are given in Appendix \ref{app2}) and leads to the representation of the effective action as an integral over the four-momentum $k_\alpha$
\begin{equation}
\label{logdet2}  
     W_E=\! - \!\int\!\! d^4x\!\!\int\!\!\frac{d^4k}{(2\pi )^4}\, e^{-k^2}\!
            \!\! \int\limits^\infty_0\!\!\frac{dt}{2t^3}\,\rho_{t,\Lambda}\,
          \mbox{tr}_I\! \left[e^{-t(M^2+A)}\right]\! ,
\end{equation}
where 
\begin{equation}
\label{A}
A = -d^2 -2ik d / \sqrt{t} +Y
\end{equation} 
is a self-adjoint operator in Hilbert space, and the summation over four-vector indices in (\ref{A}) are implicit. 

\subsection{The case of equal masses}
Before proceed with our calculations, it is appropriate to discuss the simplest case of the large-$M$ expansion, i.e., the case when the mass matrix $M$ is $M=\mbox{diag}(m,m,m)$; then $[M, A]=0$ and we have
\begin{equation}
\label{factor}
   e^{-t(M^2+A)}=e^{-tM^2}e^{-tA}=e^{-tm^2}\sum^\infty_{n=0}t^n a_n(x,x)\,.
\end{equation}
Here $a_n(x,x)$ are the Seeley - DeWitt coefficients $a_n(x,y)$ in a coincidence limit $x=y$, which depend on the background fields and their derivatives, except $a_0(x,x)=1$. Integration over four-momentum and proper-time in (\ref{logdet2}) is straightforward and we obtain a well-known result
\begin{equation}
\label{logdet3}  
     W_E= - \int\!\!\frac{d^4x}{32\pi^2}\sum_{n=0}^{\infty}
          J_{n-1}(m^2)\,\mbox{tr}_I\,a_n (x,x),
\end{equation}
where the proper-time integrals $J_n(m^2)$ are  
\begin{equation}
\label{Jnm}
   J_n(m^2)=\int\limits^\infty_0\frac{dt}{t^{2-n}}e^{-tm^2}\rho_{t,\Lambda}\, .
\end{equation}
In the case of two subtractions $\rho_{t,\Lambda}=1-(1+t\Lambda^2)e^{-t\Lambda^2}$ at the large scale $\Lambda$, one finds from (\ref{Jnm})
\begin{eqnarray}
\label{J0}
   J_0(m^2)&=&\Lambda^2-m^2\ln\left(1+\frac{\Lambda^2}{m^2}\right)\,,  \\
\label{J1}   
   J_1(m^2)&=&\ln\left(1+\frac{\Lambda^2}{m^2}\right) -\frac{\Lambda^2}{\Lambda^2+m^2}\,.
\end{eqnarray}
The choice of the regularization is closely related to the specific problem under study. Various examples of the proper-time regularization can be found in \cite{Ball:89}. The Pauli-Villars regularization we used is usually applied in the NJL model, where the cutoff $\Lambda$ is a scale of spontaneous chiral symmetry breaking.

The functions $J_n(m^2)$, for $n>1$, as $m^2$ becomes very large are asymptotically equivalent to $m^{-2(n-1)}$, that is, the expansion (\ref{logdet3}) is in inverse powers of $m^2$. For the series to converge, it is necessary not only that the mass $m$ be large, but also that the background fields change slowly over distances of the order of the fermion field Compton wavelength $1/m$. If these criteria are not met, then the production of real quark-antiquark pairs becomes essential, and the expansion is not suitable for applications. 

\subsection{The case of unequal masses}
Let us return now to the formula (\ref{logdet2}) and show how one can extend the above tool to the case $[M, A]\neq 0$. To make progress in our calculations, we use the formula (\ref{alg}) allowing to factorize the exponent with a non-commuting diagonal mass matrix $M$. Under a flavor trace, it yields, for instance, 
\begin{equation}
\mbox{tr}_f \left(e^{-tM^2}\right)=\sum_{i=1}^{3} e^{-tm_i^2}\, \mbox{tr}_f  E_i=\sum_{i=1}^3 e^{-tm_i^2}\,.
\end{equation}
The result of calculations for the remaining terms can be represented by the formula
\begin{eqnarray}
\label{c1n}
   &&\mbox{tr}_f \left(e^{-tM^2}f_n(t,A)\right) \nonumber \\
   &&= \frac{t^n}{n!} \sum_{i_1,i_2,\ldots ,i_n}^{N_f}\!\!\!\!c_{i_1i_2\ldots i_n}(t)\,
   \mbox{tr}_f(A_{i_1}A_{i_2}\ldots A_{i_n}),  
\end{eqnarray}
where $n\geq 1$ and the notation $A_i\equiv E_i A$ is used. The coefficients $c_{i_1i_2 \ldots i_n} (t)$ are totally symmetric with respect to any permutation of indices and are easily calculated. The necessary details of such calculations and useful properties of coefficients are collected in Appendix \ref{app3}. 

To ensure the fundamental cyclic property of the trace $\mbox{tr}(AB)=\mbox{tr}(BA)$, we define 
\begin{eqnarray}
\label{perm}
\mbox{tr}_f\, (A_{i_1}A_{i_2}\ldots A_{i_n})&=&\frac{1}{n}\!\!\sum_{cycl. perm.}\!\!\!\!\! A_{i_1i_2}A_{i_2i_3}\ldots A_{i_ni_1}
\nonumber \\
&\equiv& \langle A_{i_1i_2}A_{i_2i_3}\ldots A_{i_ni_1} \rangle .
\end{eqnarray}
Here we used Eqs.\,(\ref{projprop}) and (\ref{trpropEij}) to calculate the trace. The sum over a cyclic permutation of $A_{ij}$ adds nothing to the standard definition of a trace if $A$ is a matrix. However, in the case when $A$ contains open derivatives, i.e., is a differential operator, cyclic permutation in the trace may change the result. This is why it is necessary to ensure the guaranteed fulfillment of this fundamental property of the trace, which we do with the formula (\ref{perm}).

It should be emphasized that here we will restrict ourselves to considering only those terms that survive in the limit $M\to\infty$. In the case of equal masses, this approximation corresponds to considering heat coefficients $a_1$ and $a_2$. To isolate such a contribution, it is necessary to limit ourselves to the terms of order $t^2$ at most. It means that at the level of $A_i$-dependent expressions one should expand up to and including the $t^4$ order, since $A_i$ has the term $\propto 1/\sqrt t$,
 \begin{eqnarray} 
\label{sum2}
 &&\mbox{tr}_f\left(e^{-t(M^2+A)}\right)=\sum_{i=1}^3 c_i(t) -t\sum_{i=1}^3  c_i(t)\, \mbox{tr}_f\, A_i \nonumber\\ 
 &+&\frac{t^2}{2!}\sum_{i,j} c_{ij}(t) \, \mbox{tr}_f\, (A_i A_j) -\frac{t^3}{3!}\sum_{i,j,k} c_{ijk} (t)\, \mbox{tr}_f\, (A_iA_jA_k) \nonumber\\
& +&\frac{t^4}{4!}\sum_{i,j,k,l} c_{ijkl} (t)\, \mbox{tr}_f\, (A_iA_jA_kA_l)+{\cal O}(t^5)\,. 
\end{eqnarray}
Note, that the series is of the mixed type, i.e., it has also exponents depending on $t$, hidden in the coefficients $c_{i_1i_2\ldots i_n}(t)$. In the limit $m_1=m_2=m_3=m$, these coefficients shrink to $c(t)=\exp(-tm^2)$ and the mass-dependent exponent is totally factorized. This way one can recover the standard inverse mass expansion.

As it will be shown below, the term with the coefficient $c_{i_1i_2\ldots i_n}$ corresponds to the contact contribution of the Feynman diagram with $n$ internal quark lines. It means that $c_i$ can be associated with the tadpole contribution, $c_{ij}$ with the self-energy part, $c_{ijk}$ and $c_{ijkl}$ with the triangle and box contributions. The indices contain also information on the flavor content of such one-loop diagrams. For instance, $c_{123}(t)$ corresponds to the triangle originated by propagators of up, down, and strange constituent quarks.          

Now, one can substitute $A_i=E_i A$ by its expression (\ref{A}) and integrate over four-momentum $k_\alpha$ in (\ref{logdet2}) by using formula 
\begin{equation}
\int\!\! \frac{d^4k}{(2\pi )^4} e^{-k^2} k_{\alpha_1}k_{\alpha_2}\ldots k_{\alpha_{2n}}=\frac{\delta_{\alpha_1\alpha_2\ldots \alpha_{2n}}}{(4\pi)^2 2^n}\,.
\end{equation}
It is evident that the corresponding integral of an odd number of four-momentum $k_\alpha$ is zero. Totally symmetric tensor $\delta_{\alpha_1\alpha_2\ldots \alpha_{2n}}$ is determined by the recurrent relation
\begin{equation}
\delta_{\alpha_1\alpha_2\ldots \alpha_{2n}}=\sum_{i=2}^{2n} \delta_{\alpha_1}^{\alpha_i}\delta_{\alpha_2\ldots \alpha_{i-1}\alpha_{i+1}\ldots \alpha_{2n}}\,,
\end{equation}
see also Eq. (\ref{daaaa}) for details. This yields
\begin{eqnarray}
\label{HKC}
&&\int\!\!\frac{d^4k}{(2\pi)^4}\,e^{-k^2} \mbox{tr}_f\left(e^{-t(M^2+A)}\right) \nonumber \\
&&=\frac{1}{(4\pi)^2}
\left\{\sum_{i=1}^3 c_i(t)-t\sum_{i=1}^3 c_i(t)\, \mbox{tr}_f \left[E_i(Y-d^2)\right] \right. \nonumber \\
&&-t\sum_{i,j} c_{ij}(t)\,\mbox{tr}_f (E_id_\alpha E_j d_\alpha ) \nonumber \\
&&+\frac{t^2}{2}\sum_{i,j} c_{ij}(t)\,\mbox{tr}_f \left[E_i(Y-d^2)E_j(Y-d^2)\right] \nonumber \\
&&+\frac{t^2}{3}\sum_{i,j,k} c_{ijk}(t)\,\mbox{tr}_f\left[E_id_\alpha E_j d_\alpha E_k(Y-d^2)\right. \nonumber \\
&&+\left. E_id_\alpha E_j (Y-d^2) E_k  d_\alpha   +E_i (Y-d^2) E_j d_\alpha E_k d_\alpha\right] \nonumber \\
&&\left.+\frac{t^2}{3!}\sum_{i,j,k,l} c_{ijkl}(t)\delta_{\alpha\beta\gamma\delta}\,\mbox{tr}_f\left(E_id_\alpha E_j d_\beta E_k d_\gamma E_l d_\delta \right) \right\} \nonumber \\
&&+{\cal O}(t^3).
\end{eqnarray}

It remains to put this result into (\ref{logdet2}) and calculate the integrals over the proper-time. Integration turns the coefficients $c_{i_1i_2 \ldots i_n} (t)$ into functions $J_{i_1i_2 \ldots i_n}$ depending on the masses of the fermion fields and the cutoff $\Lambda$. These functions describe the leading contributions in the $1/M^2$ expansion of the corresponding one-loop Feynman diagrams, to be precise, their local parts which dominate in the limit $M\to\infty$. Here we present the result of such calculations. All necessary details can be found in Appendix \ref{app4}.

\begin{eqnarray}
\label{WE}
W_E&=&- \frac{N_c}{32\pi^2}\int d^4x \left\{-\sum_i J_0(m_i^2) \,\mbox{tr}_{Df} \left[E_i(Y-d^2) \right]  \right.\nonumber \\ 
&-&\sum_{i,j}J_0(m_i^2,m_j^2) \,\mbox{tr}_{Df}(E_id_\alpha E_j d_\alpha)  \nonumber \\
&+&\frac{1}{2}\sum_{i,j} J_{ij}\,\mbox{tr}_{Df} \left[E_i(Y-d^2)E_j(Y-d^2)\right] \nonumber \\
&+&\frac{1}{3}\sum_{i,j,k} J_{ijk} \,\mbox{tr}_{Df}\left[ E_id_\alpha E_j d_\alpha E_k(Y-d^2)\right. \nonumber \\
&+&\left. E_id_\alpha E_j (Y-d^2) E_k  d_\alpha +E_i (Y-d^2) E_j d_\alpha E_k d_\alpha \right] \nonumber \\
&+&\left.\frac{1}{6}\sum_{i,j,k,l} J_{ijkl} \delta_{\alpha\beta\gamma\delta}\,\mbox{tr}_{Df}\left[E_id_\alpha E_j d_\beta E_k d_\gamma E_l d_\delta \right] \right\}  \nonumber \\
&+&{\cal O}(1/M^2)\equiv - \frac{N_c}{32\pi^2}\!\!\int\!\! d^4x \sum_{n=0}^{\infty}\,\mbox{tr}_{D}\,b_n(x,x),
\end{eqnarray}
where the index $n$ indicates the asymptotic behavior of $b_n(x,x)$ at large masses, namely $b_n\sim M^{-2(n-2)}$. The coefficients $b_n(x, x)$ depend on the external fields and quark masses, i.e., they contain information about both the effective meson vertices and corresponding coupling constants. If all masses are equal, the dependence on $m$ is factorized in form of the integral (\ref{Jnm}) and the field-dependent part takes a standard Seeley-DeWitt form $a_n(x,x)$.

\section{Leading terms of the $1/M$ expansion}
\label{L-t-exp}

Consider the leading terms $b_1$ and $b_2$. The case $n = 0$ is of no interest because $b_0$ contains no fields and can be omitted from the effective action. The coefficients with $n\geq 3$ tend to zero in the limit of infinite masses, therefore they are small in comparison with $b_1$ and $b_2$ and, at the first stage, can be neglected. 

\subsection{Coefficient $b_1(x,x)$}    
Let us turn to the calculation of the coefficient $b_1$. It is given by a part of the expression (\ref{WE}), which is proportional to the proper-time integrals $J_0(m_i^2,m_j^2)$. With the use of Eq.\,(\ref{simplestsum}), it can be rewritten as 
\begin{eqnarray}
\label{J0part}
b_1(x,x)=&-& \sum_i J_0(m_i^2)\, \mbox{tr}_{f}\!\left[E_i(Y-d^2) + E_i d_\alpha E_i d_\alpha \right] \nonumber \\
&-&2\sum_{i<j} J_0 (m_i^2,m_j^2)\, \mbox{tr}_{f} (E_i d_\alpha E_j d_\alpha ).
\end{eqnarray}
Noting that 
\begin{equation}
\label{d2}
d^2=\partial^2 +2i\Gamma\partial +i\left(\partial \Gamma\right) -\Gamma^2,
\end{equation}
and taking into account that the action of the open derivatives in (\ref{J0part}) on the implied unit on the right hand side of Eq.\,(\ref{WE}) gives zero, we find
\begin{equation}
\mbox{tr}_{f}\!\left[ E_i(Y-d^2)\right]=\mbox{tr}_{f}\!\left[E_i(Y+\Gamma^2-i\partial \Gamma )\right].
\end{equation}
Since the last term here is a total divergence which can be omitted in the effective action, we conclude
\begin{equation}
\mbox{tr}_{f}\!\left[E_i(Y-d^2)\right]=\mbox{tr}_{f}\!\left[E_i(Y+\Gamma^2)\right]=Y_{ii}+(\Gamma^2)_{ii},
\end{equation}
where on the last stage we used Eq.\,(\ref{trpropEij}) and the fact that matrix $A$ may be written in a unique way as a finite linear combination of elements of $A$ in the bases $E_{ij}$, namely $A=\sum_{m,n}A_{mn}E_{mn}$. In the same manner, we find 
 \begin{equation}
\mbox{tr}_{f}\left(E_i d_\alpha E_j d_\alpha \right) =-\mbox{tr}_{f} (E_i \Gamma_\alpha E_j \Gamma_\alpha )=-\Gamma^\alpha_{ij}\Gamma^\alpha_{ji}.
\end{equation}

It gives for Eq.\,(\ref{J0part}) 
\begin{eqnarray}
\label{J0firstline}
b_1=&-&\sum_i J_0(m_i^2) \left[Y_{ii}+(\Gamma^2)_{ii}- \Gamma^\alpha_{ii}\Gamma^\alpha_{ii} \right] \nonumber \\
&+& 2\sum_{i<j} J_0 (m_i^2,m_j^2) \Gamma^\alpha_{ij}\Gamma^\alpha_{ji}\,.
\end{eqnarray}

Let us use now the  easily verifiable  relation 
\begin{equation}
(\Gamma^2)_{ii}- \Gamma^\alpha_{ii}\Gamma^\alpha_{ii}=\sum_{j\neq i} \Gamma^\alpha_{ij}\Gamma^\alpha_{ji}\,,
\end{equation}
to obtain finally
\begin{equation}
\label{b1}
b_1=-\sum_i J_0(m_i^2)Y_{ii}+\sum_{i<j}\Delta J_0(m_i^2,m_j^2)\Gamma^\alpha_{ij}\Gamma^\alpha_{ji}\,,
\end{equation}  
where $\Delta J_0(m_i^2,m_j^2)$ is given by Eq.\,(\ref{DJ0}). The latter integrals can be collected in the symmetric $3\times 3$ matrix $\Delta J_0$, which has zeros on the main diagonal. In particular, when $m_i = m_j = m$, all elements of $\Delta J_0$ vanish. In this specific case, the first term of (\ref{b1}) leads to the well known expression of the Seeley-DeWitt coefficient $a_1(x,x)=-Y$.

For convenience of writing the result of our calculations, along with the usual matrix multiplication, we will use the non-standard Hadamard product \cite{Styan:73}, which is the matrix of elementwise products 
\begin{equation}
(A\circ B)_{ij} =A_{ij} B_{ij}.
\end{equation}
The Hadamard product is commutative unlike regular matrix multiplication, but the distributive and associative properties are retained. It has previously been proven to be a useful tool when the mass matrices of the type (\ref{mass}) are involved \cite{Morais:17}. In terms of Hadamard product the result (\ref{b1}) can be written as 
\begin{equation}
\label{b1final}
b_1= \mbox{tr}_f\!\left[ J_0\circ (-Y) + \frac{1}{4} (\Delta J_0 \circ \Gamma^\alpha ) \Gamma^\alpha\right], 
\end{equation}
where $J_0$ is considered as a diagonal matrix with elements given by $(J_0)_{ij}=\delta_{ij} J_0(m_i^2)$. This matrix contains contributions of the Feynman one-loop diagram, known as a "tadpole". 

\subsection{Coefficient $b_2(x,x)$}  
Consider the second coefficient $b_2(x,x)$. In accordance with the general structure of the expression, we distinguish three contributions differing in the degree of $Y$
\begin{equation}
b_2=b_2^{(0)}+b_2^{(1)}+b_2^{(2)}
\end{equation}
which is explicitly indicated in the parentheses. 

\subsubsection{Quadratic part in $Y$: $b_2^{(2)}$} 
The part of $b_2$ proportional to $Y^2$ is calculated most simply
\begin{eqnarray}
\label{b2Y2}
b_2^{(2)}&=&\frac{1}{2}\sum_{i,j}J_{ij}\,\mbox{tr}_f\left(E_i YE_j Y\right)=\frac{1}{2}\sum_{i,j}J_{ij} Y_{ij} Y_{ji} \nonumber \\
&=&\frac{1}{2}\, \mbox{tr}_f\!\left[\left(J\circ Y\right)Y\right],
\end{eqnarray} 
where $J$ is a symmetric $ 3\times 3 $ matrix, whose elements $J_{ij} = J_1(m_i, m_j)$ are logarithmically divergent parts (at $\Lambda \to \infty $) of Feynman self-energy diagrams with masses of virtual particles $m_i$ and $m_j$ (see Eq.\,(\ref{J-ij})).

\subsubsection{Linear part in $Y$: $b_2^{(1)}$}  
Let us consider now the linear in $Y$ part of $b_2$, i.e., $b_2^{(1)}$ 
\begin{eqnarray} 
\label{b2(1)} 
b_2^{(1)}&=&-\frac{1}{2}\sum_{i,j} J_{ij}\,\mbox{tr}_{f}\! \left(E_id^2E_jY+E_iYE_jd^2\right) \nonumber \\
&+&\frac{1}{3}\sum_{i,j,k} J_{ijk} \,\mbox{tr}_{f}\!\left(E_id_\alpha E_j d_\alpha E_k Y\right. \nonumber \\
&+& \left. E_id_\alpha E_j Y E_k d_\alpha + E_i Y E_j d_\alpha E_k d_\alpha\right).
\end{eqnarray} 
After evaluation of traces it gives 
\begin{equation} 
b_2^{(1)}=-\frac{1}{2}\sum_{i,j} J_{ij} t_{ij}+  \sum_{i,j,k} J_{ijk} t_{ijk},
\end{equation} 
where 
\begin{equation}
\label{tijandtijk}
t_{ij}=\left\{(d^2)_{ij}, Y_{ji}\right\},  \quad t_{ijk}=\langle d^\alpha_{ij} d^\alpha_{jk} Y_{ki} \rangle .
\end{equation}

Next, let us use Eqs.\,(\ref{simplestsum}) and (\ref{3ind}) of Appendix \ref{app5}. We obtain 
\begin{eqnarray}
\label{b2se}
b_2^{(1)}&=&\sum_i J_i \left(t_{iii}-\frac{1}{2} t_{ii}\right) +\sum_{i<j} J_{ij} \left(3t^+_{iij}-t_{(ij)}\right) \nonumber \\
&+&3\sum_{i<j} R_{ij}t^-_{iij} +6J_{123}t_{(123)}.
\end{eqnarray}

To evaluate the first sum in this expression, we employ the explicit form of $t$-coefficients (\ref{tijandtijk})
\begin{eqnarray}
&& t_{iii}-\frac{1}{2} t_{ii}=\left\{ \frac{1}{3} d^\alpha_{ii} d^\alpha_{ii}-\frac{1}{2} (d^2)_{ii}, Y_{ii} \right\} +\frac{1}{3} d^\alpha_{ii} Y_{ii} d^\alpha_{ii}   \nonumber   \\
&&=-\frac{1}{6}\left[ d^\alpha_{ii}, \left[ d^\alpha_{ii}, Y_{ii}\right] \right] -\frac{1}{2}\sum_{j\neq i} \left\{ d^\alpha_{ij} d^\alpha_{ji}, Y_{ii}\right\}.
\end{eqnarray}
The first term is now computed as 
\begin{equation}
\left[ d^\alpha_{ii}, \left[ d^\alpha_{ii}, Y_{ii}\right] \right] =\partial_\alpha^2 Y_{ii}. 
\end{equation}
This is a four-divergence of a four-vector. Therefore, it can be omitted from the effective action, because it does not affect the dynamic characteristics of the effective theory. Thus, we find
\begin{equation}
\label{1coeffofsum}
t_{iii}-\frac{1}{2} t_{ii} = \sum_{j\neq i} \Gamma^\alpha_{ij} \Gamma^\alpha_{ji} Y_{ii}, 
\end{equation}
where we used the cyclic permutation of the elements to simplify the right hand side of this expression, that is not forbidden due to the trace, tr$_D$, in (\ref{WE}). In what follows, we will always use the cyclic permutation property when calculating the t-coefficients.

To evaluate the second sum in (\ref{b2se}) we need to find the difference $3t^+_{iij}-t_{(ij)}$, where $i\neq j$. Here we have 
\begin{eqnarray}
&& 3t^+_{iij}-t_{(ij)}=3\left(t_{(iij)} + t_{(jji)} \right) -\frac{1}{2}\left(t_{ij}+t_{ji} \right) \nonumber \\
&& = t_{iij}+t_{iji}+t_{jii}-\frac{1}{2} t_{ij}+(i \leftrightarrow j ) \nonumber \\
&&=\frac{1}{3}\left(\left\{d^\alpha_{ii}d^\alpha_{ij}+d^\alpha_{ij}d^\alpha_{jj}, Y_{ji}\right\}  +d^\alpha_{ij} Y_{ji} d^\alpha_{ii} + d^\alpha_{jj} Y_{ji} d^\alpha_{ij}  \right. \nonumber \\
&&+ \left.\left\{ d^\alpha_{ij}d^\alpha_{ji}, Y_{ii} \right\} + d^\alpha_{ji} Y_{ii} d^\alpha_{ij} \right) -\frac{1}{2} \{(d^2)_{ij}, Y_{ji}\}
+ (i\leftrightarrow j) \nonumber \\
&&=-\frac{1}{6}\left[\left\{d^\alpha_{ii}d^\alpha_{ij}+d^\alpha_{ij}d^\alpha_{jj}, Y_{ji}\right\}-2\left(d^\alpha_{ij} Y_{ji} d^\alpha_{ii} + d^\alpha_{jj} Y_{ji} d^\alpha_{ij} \right)\right] \nonumber \\
&&+\frac{1}{3}\left(\left\{ d^\alpha_{ij}d^\alpha_{ji}, Y_{ii} \right\} + d^\alpha_{ji} Y_{ii} d^\alpha_{ij} \right)
-\frac{1}{2} \{d^\alpha_{ik}d^\alpha_{kj}, Y_{ji}\}|_{k\neq i,j} \nonumber \\
&&+ (i\leftrightarrow j), 
\end{eqnarray}
where, on the last stage, we took into account that
\begin{equation}
\label{d^2dd}
(d^2)_{ij}=\sum_{k} d^\alpha_{ik}d^\alpha_{kj}.
\end{equation}
Noticing that expressions 
\begin{eqnarray}
&&\left\{d^\alpha_{ii}d^\alpha_{ij}+d^\alpha_{ij}d^\alpha_{jj}, Y_{ji}\right\}-2\left(d^\alpha_{ij} Y_{ji} d^\alpha_{ii} + d^\alpha_{jj} 
Y_{ji} d^\alpha_{ij} \right)\! , \nonumber \\  
&&\left\{ d^\alpha_{ij}d^\alpha_{ji}, Y_{ii} \right\} -2 d^\alpha_{ji} Y_{ii} d^\alpha_{ij} 
\end{eqnarray}
vanish under the Dirac matrix trace in (\ref{WE}), we finally obtain
\begin{equation}
\label{2coeffofsum}
3t^+_{iij}-t_{(ij)}=- \Gamma^\alpha_{ij}  \Gamma^\alpha_{ji} Y_{ii} + \Gamma^\alpha_{ik} \Gamma^\alpha_{kj} Y_{ji} + (i\leftrightarrow j),
\end{equation}
where $i \neq j\neq k$.

Since the first term in (\ref{2coeffofsum}) has the same form as (\ref{1coeffofsum}), it follows that 
\begin{eqnarray}
&&\sum_i J_i \left(t_{iii}-\frac{1}{2} t_{ii}\right) - \sum_{i<j} J_{ij}\left[\Gamma^\alpha_{ij}  \Gamma^\alpha_{ji} Y_{ii}  +(i\leftrightarrow j) \right] \nonumber \\
&&= \sum_{i,j \atop i \neq j} (J_i -J_{ij}) \Gamma^\alpha_{ij}  \Gamma^\alpha_{ji} Y_{ii}.
\end{eqnarray}

We need now to calculate the coefficient
\begin{eqnarray}
\label{3coeffofsum} 
&&t^-_{iij}=t_{(iij)} - t_{(jji)}=\frac{1}{3}\left[t_{iij}+t_{iji}+t_{jii} - (i\leftrightarrow j)  \right] \nonumber \\
&&=-\frac{1}{3}\left[\frac{i}{3} \left(\Gamma^\alpha_{ij}\! \stackrel{\leftrightarrow}{\partial_\alpha}\! Y_{ji} \right)+(\Gamma^\alpha_{ii}\Gamma^\alpha_{ij}-\Gamma^\alpha_{ij}\Gamma^\alpha_{jj})Y_{ji} \right.\nonumber \\
&&+\left.\Gamma^\alpha_{ij}\Gamma^\alpha_{ji} Y_{ii}\frac{}{}\right] -(i\leftrightarrow j),
\end{eqnarray} 
where the left-right derivative is defined in a standard way $\Gamma^\alpha\! \stackrel{\leftrightarrow} {\partial^\alpha} \! Y = \Gamma^\alpha \partial^\alpha Y - (\partial^\alpha \Gamma^\alpha) Y$. Then the third sum in (\ref{b2se}) is  
\begin{eqnarray} 
&&3\sum_{i<j} R_{ij}t^-_{iij}= - \sum_{i,j} R_{ij} \left[\frac{i}{3} \left(\Gamma^\alpha_{ij}\! \stackrel{\leftrightarrow}{\partial_\alpha}\! Y_{ji} \right) \right. \nonumber \\
&&+(\Gamma^\alpha_{ii}\Gamma^\alpha_{ij}-\Gamma^\alpha_{ij}\Gamma^\alpha_{jj})Y_{ji} +\left.\Gamma^\alpha_{ij}\Gamma^\alpha_{ji} Y_{ii}\frac{}{}\right].
\end{eqnarray} 
 
The coefficients $t_{(123)}$ are required for the evaluation of the last term in (\ref{b2se}). They can be obtained with the aid of the easily verifiable relations,  
\begin{equation}
t_{ijk}|_{i\neq j\neq k}= -\Gamma^\alpha_{ij}\Gamma^\alpha_{jk} Y_{ki}.
\end{equation} 
As a consequence, we have
\begin{equation}
t_{(123)}=\frac{1}{6}\sum_{i\neq j\neq k} t_{ijk}=-\frac{1}{6} \sum_{i\neq j\neq k} \Gamma^\alpha_{ij}\Gamma^\alpha_{jk} Y_{ki}. 
\end{equation} 
 
If we take into account the existence of similar term in the coefficient (\ref{2coeffofsum}), we obtain
\begin{eqnarray}
&&6J_{123}t_{(123)}+\sum_{i<j \atop i \neq j \neq k} J_{ij} \left[ \Gamma^\alpha_{ik} \Gamma^\alpha_{kj} Y_{ji} + (i\leftrightarrow j)\right] \nonumber \\
&&= \sum_{i\neq j\neq k} (J_{ij}-J_{ijk}) \Gamma^\alpha_{ij}\Gamma^\alpha_{jk} Y_{ki}.
\end{eqnarray} 
 
Thus, the final result for $b_2^{(1)}$ is 
\begin{eqnarray} 
\label{b21final}
b_2^{(1)} &=& \sum_{i\neq j\neq k} \left[ (J_i -J_{ij}) \Gamma^\alpha_{ij}  \Gamma^\alpha_{ji} Y_{ii} 
+  (J_{ij}-J_{ijk}) \Gamma^\alpha_{ij}\Gamma^\alpha_{jk} Y_{ki}\right] \nonumber \\
&-& \sum_{i,j} R_{ij} \left[\frac{i}{3} \left(\Gamma^\alpha_{ij}\! \stackrel{\leftrightarrow}{\partial^\alpha}\! Y_{ji} \right) +(L_-^{\alpha\alpha})_{ij}Y_{ji} +\Gamma^\alpha_{ij}\Gamma^\alpha_{ji} Y_{ii}\right]  \nonumber \\
&=& \mbox{tr}_f \left[ NY-\frac{i}{3}(R\circ \Gamma^\alpha )\! \stackrel{\leftrightarrow}{\partial^\alpha}\! Y \right],
\end{eqnarray}
where we introduce the notation
\begin{equation}
\left(L^{\alpha\beta}_\pm\right)_{ij}=\left(\Gamma^\alpha_{ii}\pm \Gamma^\alpha_{jj}\right)\Gamma^\beta_{ij}.
\end{equation}
Note that elements of the matrix $\Gamma^\alpha_{ij}$ commute with each other. In the last line of Eq.\,(\ref{b21final}), $R$ and $N$ are the $3\times 3$ matrices with elements  
\begin{eqnarray}
 R_{ij} &=& \frac{1}{2} \left(  J_{iij}-J_{jji}  \right),   \nonumber \\
 N_{ii} &=& \sum_{j\neq i} \left(J_{i}-J_{ij}-R_{ij}\right)\Gamma^\alpha_{ij} \Gamma^\alpha_{ji}\,,  \\
 N_{ij} &=& \left[(J_{ij}-J_{ijk})\Gamma^\alpha_{ik} \Gamma^\alpha_{kj} -R_{ij}(L_-^{\alpha\alpha})_{ij} 
\right]_{\mid i\neq j\neq k}. \nonumber 
\end{eqnarray} 
In the limit of equal masses, the coefficient $b_2^{(1)}$ vanishes, that is, it contains only contributions associated with an explicit violation of chiral symmetry. Obviously, without this term, the description of the manifest flavor symmetry breaking would be incomplete.

\subsubsection{Independent on $Y$ part: $b_2^{(0)}$}  
The part of $b_2$ which does not depend on $Y$ is  
\begin{eqnarray}
\label{b20sums}
b_2^{(0)}&=& \frac{1}{2}\sum_{i,j} J_{ij} (d^2)_{ij} (d^2)_{ji} -\sum_{i,j,k} J_{ijk}\hat t_{ijk}  \nonumber \\
              &+&\frac{1}{6}\sum_{i,j,k,l} J_{ijkl} t_{ijkl}\,. 
\end{eqnarray} 
Here the field dependent t-coefficients are given by the following formulas. 
\begin{equation}
\label{coeff-hatt}
\hat t_{ijk} =\langle d^\alpha_{ij} d^\alpha_{jk} (d^2)_{ki} \rangle , \quad
t_{ijkl}= \delta_{\alpha\beta\gamma\delta} \langle d^\alpha_{ij} d^\beta_{jk} d^\gamma_{kl} d^\delta_{li}\rangle . 
\end{equation} 

Before of calculating (\ref{b20sums}) directly, we shall first handle the case of equal masses $m_u=m_d=m_s=m$, which can easily be obtained from (\ref{b20sums}) after factorizing the common factor $J_1(m^2)=J_{ii}=J_{iii}=J_{iiii}$   
\begin{eqnarray}
\label{limcaseb20}
&&\frac{b_2^{(0)}|_{m}}{J_1(m^2)}= \mbox{tr}_f\left[ \frac{1}{2} d^2 d^2 -\frac{1}{3} \left(d^\alpha d^\alpha d^2 +d^\alpha d^2 d^\alpha    \right.  \right. \nonumber \\
&& + \left. d^2 d^\alpha d^\alpha \right) +\frac{1}{6} \left( d^\alpha d^\alpha d^\beta d^\beta + d^\alpha d^\beta d^\alpha d^\beta + d^\alpha d^\beta d^\beta d^\alpha \right)\left.\frac{}{}\!\!\!\right] \nonumber \\
&&=\frac{1}{6} \mbox{tr}_f\left(d^\alpha d^\beta d^\alpha d^\beta - d^\alpha d^\beta d^\beta d^\alpha \right)=\frac{1}{12}\mbox{tr}_f\! \left[d^\alpha , d^\beta \right]^2  \nonumber \\
&&=-\frac{1}{12} \mbox{tr}_f\! \left(\Gamma^{\alpha\beta} \Gamma^{\alpha\beta}\right),
\end{eqnarray}
where we used Eq.(\ref{d^2dd}), and 
 \begin{equation}
 \label{GAlBe}
 \Gamma^{\alpha\beta}=\partial^\alpha\Gamma^\beta -\partial^\beta\Gamma^\alpha +i[\Gamma^\alpha, \Gamma^\beta ].
 \end{equation}  

To obtain the exact result for (\ref{b20sums}) it is convenient to consider separately the cases when the sum runs only over the same index values, and then over two and three different values of the flavors
\begin{equation}
b_2^{(0)}=\beta_2^{(1)}+\beta_2^{(2)}+\beta_2^{(3)}.
\end{equation}

The part of $b_2^{(0)}$ that corresponds to a contribution with coincident indices is   
\begin{equation}
\label{beta1}
\beta_2^{(1)}=\sum_i J_i \left[ \frac{1}{2} (d^2)_{ii} (d^2)_{ii} -\hat t_{iii} + \frac{1}{6} t_{iiii}\right].
\end{equation}
Introducing the notation
\begin{equation}
Q_{ii} =d^\alpha_{ii} d^\alpha_{ii} -(d^2)_{ii}=(\Gamma^2)_{ii} -\Gamma^\alpha_{ii} \Gamma^\alpha_{ii}\,,
\end{equation}
we can rewrite (\ref{beta1}) as
$$ 
\beta_2^{(1)}=\sum_i J_i \left\{ \frac{1}{2} Q_{ii}^2 -\frac{1}{6} [d^\alpha_{ii}, [d^\alpha_{ii}, Q_{ii}]]+\frac{1}{12} [d^\alpha_{ii}, d^\beta_{ii}]^2  \right\}.
$$ 

Now we benefit from the following easily established relations
\begin{eqnarray}
&&  [d^\alpha_{ii}, [d^\alpha_{ii}, Q_{ii}]]=\partial^2 Q_{ii}, \\
&&  [d^\alpha_{ii}, d^\beta_{ii}] = i\left(\partial^\alpha \Gamma^\beta_{ii} - \partial^\beta \Gamma^\alpha_{ii}  \right)\equiv i F^{\alpha\beta}_{ii} .
\end{eqnarray}
That gives finally 
\begin{equation}
\label{b2-1}
\beta_2^{(1)}=\sum_i J_i \left\{ \frac{1}{2} Q_{ii}^2 -\frac{1}{12} (F^{\alpha\beta}_{ii})^2  \right\}.
\end{equation}
To obtain this expression, we, as before, discarded a full divergence of a four-vector $\partial^2 Q_{ii}$.

Now consider the terms of the sum (\ref{b20sums}) with two unequal indices. These comprise of the remaining terms of the first sum and the corresponding contributions from the second and third sums
\begin{eqnarray}
\label{beta2-2}
\beta_2^{(2)}&=& \frac{1}{2} \sum_{i<j} J_{ij} \{(d^2)_{ij},  (d^2)_{ji} \} -3 \sum_{i<j}\left(J_{ij} \hat t^+_{iij} +R_{ij}  \hat t^-_{iij}  \right) \nonumber \\
&+&\sum_{i < j} \left[\frac{2}{3}\left(J_{ij}+\frac{1}{2} S_{ij}  \right) t^+_{iiij} +R_{ij} t^-_{iiij}\right] \nonumber \\
&+&\sum_{i<j} \left(J_{ij}-S_{ij}\right) t_{(iijj)}\,,
\end{eqnarray}
where $S_{ij}$ is given by Eq.\,(\ref{DJ1}).

To evaluate the first sum, we note that 
\begin{eqnarray}
\label{d2d2}
(d^2)_{ij} (d^2)_{ji} &=& (\partial\Gamma_{ij})(\partial \Gamma_{ji}) +  (\Gamma^2)_{ij}  (\Gamma^2)_{ji} \nonumber \\
&+& i(\partial \Gamma_{ij})(\Gamma^2)_{ji} - i (\Gamma^2)_{ij}  (\partial \Gamma_{ji}).
\end{eqnarray}
Here we have omitted a full divergence of a four-vector, and took into account that the action of the derivative that reached the right end of the expression gives zero. In the anticommutator $\{(d^2)_{ij},  (d^2)_{ji} \}$, only the symmetric part of Eq.\,(\ref{d2d2}) contributes to the sum. Therefore, the second line of (\ref{d2d2}) has no influence on the result, and we obtain   
\begin{eqnarray}
&&\frac{1}{2}\sum_{i<j} J_{ij} \{(d^2)_{ij}, (d^2)_{ji} \} \nonumber \\
&=& \sum_{i<j} J_{ij} \left[(\partial \Gamma_{ij})(\partial \Gamma_{ji}) +  (\Gamma^2)_{ij}  (\Gamma^2)_{ji}\right].
\end{eqnarray}
Recall again that elements of the matrix $\Gamma$ commute with each other.

Combining 
\begin{equation}
\label{t-coeff-4}
\frac{2}{3} t^+_{iiij}+t_{(iijj)} = \frac{2}{3}\left(  t_{iiij}+t_{jjji}+t_{iijj} \right)+\frac{1}{3} t_{ijij}\,,
\end{equation}
where the properties of t-coefficients (\ref{t3-1})-(\ref{t1-1-1-1}) have been used, we learn that this sum does not contain the terms with one derivative. Indeed,
\begin{eqnarray*}
&&t_{iiij}+t_{jjji}+t_{iijj} = (\partial\Gamma_{ij})(\partial\Gamma_{ji})+\frac{1}{2} (\partial^\alpha\Gamma^\beta_{ij})(\partial^\alpha\Gamma^\beta_{ji}) \nonumber \\
&&+\delta_{\alpha\beta\gamma\sigma} \left(\Gamma^\alpha_{ii}\Gamma^\beta_{ii}\Gamma^\gamma_{ij}\Gamma^\sigma_{ji}+\Gamma^\alpha_{jj}\Gamma^\beta_{jj}\Gamma^\gamma_{ji}\Gamma^\sigma_{ij} +\Gamma^\alpha_{ii}\Gamma^\beta_{ij}\Gamma^\gamma_{jj}\Gamma^\sigma_{ji}  \right). 
\end{eqnarray*}
We also know from (\ref{t_ijkl}) that
\begin{equation}
t_{ijij}=\delta_{\alpha\beta\gamma\sigma} \Gamma^\alpha_{ij}\Gamma^\beta_{ji}\Gamma^\gamma_{ij}\Gamma^\sigma_{ji}.
\end{equation}
Thus, only the coefficient $\hat t^+_{iij}$ contains terms with one derivative. The expression for this coefficient is given in Appendix \ref{app6} (see Eq.\,(\ref{hatt+iij})).

Before we turn to the full expression for the coefficient $\beta_2^{(2)}$ let us first collect terms with two derivatives in $\beta_2^{(2)}$ 
\begin{eqnarray}
\label{beta22}
\beta_2^{(2)}|_{\partial^2}&=&\frac{1}{3} \sum_{i<j} J_{ij}  \left[ (\partial\Gamma_{ij})(\partial\Gamma_{ji})-(\partial^\alpha\Gamma^\beta_{ij})(\partial^\alpha\Gamma^\beta_{ji}) \right] \nonumber \\
&=&-\frac{1}{6} \sum_{i<j} J_{ij} F^{\alpha\beta}_{ij} F^{\alpha\beta}_{ji}.
\end{eqnarray} 

Combined together with similar terms in $\beta_2^{(1)}$ they give the kinetic term of spin-1 fields
\begin{eqnarray}
\label{b20twoder}
&&\left.\left( \beta_2^{(1)}+\beta_2^{(2)}\right)\right|_{\partial^2}=-\frac{1}{12}\, \mbox{tr}_f \left[ F^{\alpha\beta} (J\circ F^{\alpha\beta})\right] \nonumber \\
&&=-\frac{1}{12}\, \mbox{tr}_f \left[ \Gamma^{\alpha\beta} (J\circ \Gamma^{\alpha\beta})\right] 
+ \frac{i}{3}\, \mbox{tr}_f \left[ F^{\alpha\beta} (J\circ \Gamma^{\alpha}\Gamma^\beta )\right] \nonumber \\
&&-\frac{1}{6}\, \mbox{tr}_f \left\{ [\Gamma^\alpha , \Gamma^\beta ]  (J\circ \Gamma^{\alpha}\Gamma^\beta )    \right\}.
\end{eqnarray}
In the case of equal masses, the first term coincides with the standard result (\ref{limcaseb20}). It means that the last two terms of (\ref{b20twoder}) after combining them with other contributions to $b_2^{(0)}$ should vanish when $m_u=m_d=m_s=m$. 

To see this we need to consider the contribution $\beta_2^{(3)}$. On referring to Eqs.\,(\ref{b20sums}) and (\ref{4in}), we find
\begin{equation}
\label{beta2(3)}
\beta_2^{(3)}=2\!\!\sum_{i\neq j\neq k \atop j<k} J_{iijk} t_{(iijk)} -6J_{123}\hat t_{(123)}.
\end{equation}

Using the explicit expression for the coefficients $t_{(iijk)}$ (\ref{t_(iijk)}) and Eq.\,(\ref{symJiijk}), one can represent the first sum as
\begin{eqnarray*}
&&2\!\!\sum_{i\neq j\neq k \atop j<k}\!\! J_{iijk} t_{(iijk)}=\sum_{i\neq j\neq k}\!\! J_{iijk} t_{(iijk)} \nonumber \\
&&=\frac{1}{3}\delta_{\alpha\beta\gamma\sigma}\!\! \sum_{i\neq j\neq k} \left[\frac{i}{2} (J_{jjik}-J_{iijk})(\partial^\alpha \Gamma^\beta_{ij} )\Gamma^\gamma_{jk}\Gamma^\sigma_{ki}\right.\nonumber \\
&&\left.+ J_{iijk}\left(\Gamma^\alpha_{ii} \Gamma^\beta_{ij} \Gamma^\gamma_{jk} \Gamma^\sigma_{ki}  + \Gamma^\alpha_{ii} \Gamma^\beta_{ik} \Gamma^\gamma_{kj} \Gamma^\sigma_{ji} 
+\Gamma^\alpha_{ij} \Gamma^\beta_{ji} \Gamma^\gamma_{ik} \Gamma^\sigma_{ki}\right)\right].
\end{eqnarray*}
If we introduce the antisymmetric matrix $K$ with elements 
\begin{equation}
\label{Kij}
K_{ij}=(J_{jjik}-J_{iijk})_{\mid i\neq j\neq k} \,,
\end{equation}
then the first part of the expression can be presented in the index-free form
\begin{equation}
\label{Eab}
\frac{i}{6} \delta_{\alpha\beta\gamma\sigma} \mbox{tr}_f\left[\left(K\circ \partial^\alpha\Gamma^\beta \right) E^{\gamma\sigma}\right],
\end{equation}
where the off-diagonal elements of matrix $E^{\gamma\sigma}$ are  
\begin{equation}
E^{\gamma\sigma}_{ij}=\Gamma^{\gamma}_{ik}\Gamma^{\sigma}_{kj}, \,\, (i\neq j\neq k). 
\end{equation}
Due to Eq.\,(\ref{Kij}), the diagonal elements of the matrix $E^{\gamma\sigma}$ do not contribute to (\ref{Eab}).

It is also clear from (\ref{Kij}) that terms with one derivative (\ref{Eab}) vanish in the limit of equal masses. Therefore, they cannot cancel the unwanted second term in (\ref{b20twoder}). 

Since the coefficient $\hat t_{(123)}$ is known from Eq.\,(\ref{hatttotsym}), we can give a complete expression for $\beta_2^{(3)}$
\begin{eqnarray}
\label{b2-3}
&&\beta_2^{(3)} =\frac{1}{3}\delta_{\alpha\beta\gamma\sigma}\!\! \sum_{i\neq j\neq k} \left[\frac{i}{2} (J_{jjik}-J_{iijk})(\partial^\alpha \Gamma^\beta_{ij} )\Gamma^\gamma_{jk}\Gamma^\sigma_{ki}\right.\nonumber \\
&&\left.+ J_{iijk}\left(\Gamma^\alpha_{ii} \Gamma^\beta_{ij} \Gamma^\gamma_{jk} \Gamma^\sigma_{ki}  + \Gamma^\alpha_{ii} \Gamma^\beta_{ik} \Gamma^\gamma_{kj} \Gamma^\sigma_{ji} 
+\Gamma^\alpha_{ij} \Gamma^\beta_{ji} \Gamma^\gamma_{ik} \Gamma^\sigma_{ki}\right)\right] \nonumber \\
&&-J_{123}\sum_{i \neq j \neq k} \left[\frac{i}{3} \Gamma^\alpha_{ij}\Gamma^\beta_{jk} F^{\alpha\beta}_{ki} + \Gamma^\alpha_{ij}\Gamma^\alpha_{jk} (\Gamma^2)_{ki} \right].
\end{eqnarray}

Now it is necessary to make sure that the terms linear in $F^{\alpha\beta}$ vanish in the case of equal fermion masses. To this end let us consider the corresponding contributions to $b_2^{(0)}$ from $\hat t^+_{(iij)}$ in (\ref{beta2-2}), from the second term of (\ref{b20twoder}) and from $\hat t_{(123)}$ of Eq.\,(\ref{beta2(3)}). Combined together they give
\begin{eqnarray} 
&&\frac{i}{3}\sum_{i \neq j \neq k} \left[(J_i-J_{ij}) \Gamma^\alpha_{ij} \Gamma^\beta_{ji} F^{\alpha\beta}_{ii} 
+(J_{ij}-J_{123})  \Gamma^\alpha_{ij} \Gamma^\beta_{jk} F^{\alpha\beta}_{ki}    \right] \nonumber \\
&&=\frac{i}{3} \mbox{tr}_f \left(H^{\alpha\beta} F^{\alpha\beta}  \right),
\end{eqnarray}
where the flavor matrix $H^{\alpha\beta}$ has elements
\begin{eqnarray}
\label{Hab}
&&H^{\alpha\beta}_{ii}=\sum_{j\neq i} (J_i-J_{ij})\Gamma^\alpha_{ij}\Gamma^\beta_{ji}, \nonumber \\
&&(H^{\alpha\beta}_{ij})|_{i \neq j}=(J_{ij}-J_{123}) E^{\alpha\beta}_{ij}. 
\end{eqnarray}
One can see that all elements of this matrix vanish in the limit of equal masses. 

Our next step is to demonstrate the fulfilment of the limiting condition for terms of the fourth order in $\Gamma$. As we already know these terms should also vanish for equal-mass case.  

In the first step, we transform the following expression
\begin{eqnarray*}
&& \frac{1}{2}\sum_i J_i Q_{ii}^2 -J_{123}\!\!\! \sum_{i\neq j\neq k}\!\! \Gamma^\alpha_{ij}\Gamma^\alpha_{jk} (\Gamma^2)_{ki} +\sum_{i<j} J_{ij}(\Gamma^2)_{ij} (\Gamma^2)_{ji}   \nonumber \\
&&-\sum_{i<j} J_{ij}\left[ \Gamma^\alpha_{ji} (\Gamma^\alpha_{ii}+\Gamma^\alpha_{jj})(\Gamma^2)_{ij} +\Gamma^\alpha_{ij} \Gamma^\alpha_{ji} (\Gamma^2)_{ii} + (i\leftrightarrow j)   \right].
\end{eqnarray*}
Here the first sum is taken from $\beta_2^{(1)}$ (see Eq.\,(\ref{b2-1})), the second one from $\beta_2^{(3)}$ (the last term in Eq.\,(\ref{b2-3})), and the third and fourth sums are contributions from $\beta_2^{(2)}$ (from the first and $\hat t^+_{iij}$ terms of (\ref{beta2-2}) correspondingly).

This expression can be made into a somewhat more simple form
\begin{equation}
\label{promex}
\frac{1}{2}\mbox{tr}_f \left[ (H^{\alpha\alpha}-\hat H^{\alpha\alpha})\Gamma^2\right]+\frac{1}{2}\sum_{i} J_i (\Gamma_{ii}\Gamma_{ii})^2,
\end{equation}
where the matrix $H^{\alpha\alpha}$ is a particular case of (\ref{Hab}), and the elements of the matrix $\hat H^{\alpha\alpha}$ are
\begin{eqnarray}
&& \hat H_{ii}^{\alpha\alpha} =\sum_j J_{ij} \Gamma^\alpha_{ij}  \Gamma^\alpha_{ji}\,, \nonumber \\
&& \hat H_{ij}^{\alpha\alpha}|_{i\neq j}=J_{ij}\Gamma^\alpha_{ij} \left(\Gamma^\alpha_{ii}+ \Gamma^\alpha_{jj}\right) +J_{123} E^{\alpha\alpha}_{ij}.
\end{eqnarray}  

Under the flavor trace in (\ref{promex}), we explicitly singled out the contribution $H^{\alpha\alpha}$ that vanishes in the case of equal masses. The last term in (\ref{promex}) is completely canceled by the corresponding contribution from $\hat H^{\alpha\alpha}$. Indeed, it is straightforward to see
\begin{eqnarray}
\label{1prom-etap}
&&-\frac{1}{2}\mbox{tr}_f \left(\hat H^{\alpha\alpha} \Gamma^2\right)+\frac{1}{2}\sum_{i} J_i (\Gamma_{ii}\Gamma_{ii})^2  \nonumber \\
&& = -\frac{1}{2} \sum_{i\neq j}  \Gamma^\alpha_{ij} \Gamma^\alpha_{ji} \left[ J_i (\Gamma_{ii} \Gamma_{ii})
+ J_{ij}(\Gamma^2)_{ii}\right] \nonumber \\
&&\ \ \  -\frac{1}{2} \sum_{i\neq j}  \hat H^{\alpha\alpha}_{ij} (\Gamma^2)_{ji}\,.
\end{eqnarray}

At the second stage, we collect three contributions: the part of $\beta_2^{(2)}$ associated with the coefficient (\ref{t-coeff-4})
\begin{equation}
\label{secterm}
\frac{1}{6}\delta_{\alpha\beta\gamma\sigma} \sum_{i\neq j} J_{ij} \Gamma^\alpha_{ij}\Gamma^\beta_{ji} \left( \Gamma^\gamma_{ij}\Gamma^\sigma_{ji} +2 \Gamma^\gamma_{ii}\Gamma^\sigma_{jj} +4 \Gamma^\gamma_{ii}\Gamma^\sigma_{ii}\right),
\end{equation} 
the last term in Eq.(\ref{b20twoder})
\begin{equation}
\label{thirdterm}
-\frac{1}{12}\sum_i J_i [\Gamma^\alpha, \Gamma^\beta]^2_{ii} -\frac{1}{6} \sum_{i<j} J_{ij} [\Gamma^\alpha, \Gamma^\beta]_{ij} [\Gamma^\alpha, \Gamma^\beta]_{ji}\,,  
\end{equation}
and the $\Gamma^4$-part of $\beta_2^{(3)}$ from Eq.\,(\ref{b2-3})
\begin{eqnarray}
\label{fourthterm}
&&\frac{1}{3}\delta_{\alpha\beta\gamma\sigma}\!\!\sum_{i\neq j\neq k}\!\! J_{iijk} \left( \Gamma^\alpha_{ii}   \Gamma^\beta_{ij}  \Gamma^\gamma_{jk}   \Gamma^\sigma_{ki} + \Gamma^\alpha_{ii}   \Gamma^\beta_{ik}  \Gamma^\gamma_{kj}   \Gamma^\sigma_{ji} \right. \nonumber \\
&&\ \ \ \ \ \ \ \ \ \ \ \ \ \ \ \ \ \ \ \    +\left. \Gamma^\alpha_{ij}  \Gamma^\beta_{ji} \Gamma^\gamma_{ik}  \Gamma^\sigma_{ki}  \right).
\end{eqnarray}
Together with the terms just considered in (\ref{1prom-etap}), they should lead to a result that vanishes in the limit of equal masses. 

Indeed, it is evident from the direct calculation of these terms that the summation gives thirty three independent combinations of $\Gamma$'s, which can be collected in the following sums 
\begin{widetext}
\begin{eqnarray}
\label{G33}
&&\Omega^{(0)}_1 = (\ref{promex})+(\ref{secterm})+(\ref{thirdterm})+(\ref{fourthterm}) = 
\sum_{i<j} \left\{
\left[\frac{2}{3} (J_{i}+J_{j})-\frac{4}{3} J_{ij}\right]
\left(\Gamma_{ij}\Gamma_{ji}\right)^2
+\frac{1}{6} (2J_{ij} -J_{i}-J_{j})
\left(\Gamma_{ij}\Gamma_{ij}\right)\left(\Gamma_{ji}\Gamma_{ji}\right) \right\} 
\nonumber \\
&&+\sum_{i\neq j\neq k, \atop j<k} \!\! \left\{
\left[\frac{1}{3}\, (J_{jk}+2J_{iijk})+J_{i}-J_{ij}-J_{ik}\right]
\left(\Gamma_{ij}\Gamma_{ji}\right)\left(\Gamma_{ik}\Gamma_{ki}\right)
+\left[\frac{1}{3}\, (J_{i}+2J_{iijk})+J_{jk}-2J_{123}\right] 
\left(\Gamma_{ji}\Gamma_{ik}\right)\left(\Gamma_{ki}\Gamma_{ij}\right) 
\right. \nonumber \\
&&+\frac{1}{3}(2J_{iijk}-J_{ij}-J_{ik}) 
\left[ \left(\Gamma_{ii}\Gamma_{jk}\right)\left(\Gamma_{ij}\Gamma_{ki}\right) + 
        \left(\Gamma_{ik}\Gamma_{ji}\right)\left(\Gamma_{kj}\Gamma_{ii}\right)
\right] 
+\left. \frac{1}{3} (2J_{iijk}-J_{i}-J_{jk})
\left(\Gamma_{ij}\Gamma_{ik}\right)\left(\Gamma_{ki}\Gamma_{ji}\right) 
\right\} \nonumber \\
&&+ \!\! \sum_{i\neq j\neq k}\left[
\frac{1}{3} (J_{ik}+2J_{iijk})-J_{123}\right] 
\left[\left(\Gamma_{ii}\Gamma_{ij}\right)\left(\Gamma_{jk}\Gamma_{ki}\right) + 
\left(\Gamma_{ik}\Gamma_{kj}\right)\left(\Gamma_{ji}\Gamma_{ii}\right)\right].  
\end{eqnarray}
\end{widetext}
Here, the expression in the parentheses $\left(\Gamma_{ij}\Gamma_{ji}\right)$ is understood to be summed over alpha $\Gamma^\alpha_{ij}\Gamma^\alpha_{ji}$. In the case of equal quark masses (\ref{G33}) is zero, since all $J$-dependent factors vanish.
 
It remains to calculate the sums containing the coefficients $R_{ij}$ and $S_{ij}$ in (\ref{beta2-2}). For that we need to know the coefficients $3\hat t^-_{iij}-t^-_{iiij}$ and $t^+_{iiij} -3t_{(iijj)}$. 

From Eqs.\,(\ref{coeff-t-iij}) and (\ref{coeff-t-pm-iiij}) one finds 
\begin{eqnarray}
&&3\hat t^-_{iij}-t^-_{iiij} = \frac{i}{12}\left[8(\partial\Gamma_{ji}) (\Gamma^2)_{ij} +F^{\alpha\beta}_{ii} \Gamma^\alpha_{ij}\Gamma^\beta_{ji} \right.  \nonumber \\
&&\left. +(F^{\alpha\beta}_{ij} +9\partial^\alpha\Gamma^\beta_{ij})L^{\beta\alpha}_{+ji}\right] +L^{\alpha\alpha}_{-ij} (\Gamma_{jk}\Gamma_{ki}) \nonumber \\
&&+(\Gamma_{ij}\Gamma_{ji})(\Gamma_{ik}\Gamma_{ki}) -(i\leftrightarrow j),
\end{eqnarray} 
where it is assumed that ${k\neq i\neq j}$. Then, due to the antisymmetry of $R_{ij}$, we find
\begin{eqnarray}
&&\sum_{i<j} R_{ji} (3\hat t^-_{iij}-t^-_{iiij})= 
\sum_{i\neq j}R_{ij} \left\{ \frac{i}{12}\left[ 8(\partial\Gamma_{ij}) (\Gamma^2)_{ji} \right.\right. \nonumber\\
&&+\left. \Gamma^\alpha_{ji}\Gamma^\beta_{ij}F^{\alpha\beta}_{jj}+(F^{\alpha\beta}_{ji} +9\partial^\alpha\Gamma^\beta_{ji})L^{\beta\alpha}_{+ij}\right] \nonumber \\
&&+\left. L^{\alpha\alpha}_{-ji} (\Gamma_{ik}\Gamma_{kj})+(\Gamma_{ji}\Gamma_{ij})(\Gamma_{jk}\Gamma_{kj})\right\}.
\end{eqnarray} 
 
Now, from Eqs.\,(\ref{coeff-t-pm-iiij}) and (\ref{coeff-t-(iijj)}) one obtains  
\begin{eqnarray}
&& t^+_{iiij} -3t_{(iijj)}= \frac{1}{4}F^{\alpha\beta}_{ij}  F^{\alpha\beta}_{ji} +\frac{3}{2}(\partial\Gamma_{ij})(\partial\Gamma_{ji}) \nonumber \\
&& -\frac{3i}{4}\!\left(\Gamma^\alpha_{ii}-\Gamma^\alpha_{jj}\right)\! \!
\left[ (\Gamma^\alpha_{ji}\! \stackrel{\leftrightarrow}{\partial^\beta}\! \Gamma^\beta_{ij})+(\Gamma^\beta_{ji}\! \stackrel{\leftrightarrow}{\partial^\alpha}\! \Gamma^\beta_{ij} ) + (\Gamma^\beta_{ji}\! \stackrel{\leftrightarrow}{\partial^\beta}\! \Gamma^\alpha_{ij} ) \right] \nonumber\\
&&+\delta_{\alpha\beta\gamma\sigma}\left(\Gamma^\alpha_{ii}\Gamma^\beta_{ii}\Gamma^\gamma_{ij}\Gamma^\sigma_{ji}    +\Gamma^\alpha_{jj}\Gamma^\beta_{jj}\Gamma^\gamma_{ji}\Gamma^\sigma_{ij}
-\Gamma^\alpha_{ij}\Gamma^\beta_{ji}\Gamma^\gamma_{ij}\Gamma^\sigma_{ji}\right. \nonumber \\
&&\left. -2\Gamma^\alpha_{ii}\Gamma^\beta_{ij}\Gamma^\gamma_{jj}\Gamma^\sigma_{ji}\right).
\end{eqnarray} 
By virtue of the symmetry $S_{ij} =S_{ji}$, the required sum is  
\begin{eqnarray}
&&\sum_{i<j} S_{ij} \left(\frac{1}{3} t^+_{iiij} - t_{(iijj)}\right) = \frac{1}{3} \sum_{i \neq j} S_{ij}   \nonumber \\
&&\!\!\!\!\!\times\!\left\{ \frac{1}{8}F^{\alpha\beta}_{ij}  F^{\alpha\beta}_{ji} +\frac{3}{4}(\partial\Gamma_{ij})(\partial\Gamma_{ji}) 
-\frac{3i}{8}\!\left(\Gamma^\alpha_{ii}-\Gamma^\alpha_{jj}\right)\right. \nonumber \\
&& \!\!\!\!\!\times\! \left[ (\Gamma^\alpha_{ji}\! \stackrel{\leftrightarrow}{\partial^\beta}\! \Gamma^\beta_{ij})+(\Gamma^\beta_{ji}\! \stackrel{\leftrightarrow}{\partial^\alpha}\! \Gamma^\beta_{ij} ) + (\Gamma^\beta_{ji}\! \stackrel{\leftrightarrow}{\partial^\beta}\! \Gamma^\alpha_{ij} ) \right] +\delta_{\alpha\beta\gamma\sigma} \nonumber \\
&&\left. \!\!\!\!\!\times\!\left(\Gamma^\alpha_{ii}\Gamma^\beta_{ii}\Gamma^\gamma_{ij}\Gamma^\sigma_{ji}-\Gamma^\alpha_{ii}\Gamma^\beta_{ij}\Gamma^\gamma_{jj}\Gamma^\sigma_{ji}
-\frac{1}{2}\Gamma^\alpha_{ij}\Gamma^\beta_{ji}\Gamma^\gamma_{ij}\Gamma^\sigma_{ji}\right)\!\!\right\}\!\!.
\end{eqnarray} 
Now we have all the base ingredients that remain to be assembled into a single whole.
 
\subsubsection{Full result for $b_2(x,x)$}  
Here we summarize the above results collecting them in a compact form convenient for further applications. Namely, the result obtained for $b_2(x,x)$ can be represented as 
\begin{equation}
\label{b2}
b_2=\frac{1}{2} \mbox{tr}_f\! \left[ Y\left(J\circ Y\right)-\frac{1}{6} \Gamma^{\alpha\beta} \left(J\circ \Gamma^{\alpha\beta} \right)\right]+\Delta b_2.
\end{equation}
In the limit of equal masses, the first two terms in (\ref{b2}) yield the well-known result \cite{Ball:89}, and the third one vanishes, that is, it contains only contributions associated with an explicit violation of chiral symmetry. Together with the second term of Eq.\,(\ref{b1}), this is the main result of our work. 

To write down the full expression for $\Delta b_2$, we distinguish four different contributions 
\begin{equation}
\label{db2}
\Delta b_2=b_2^{(1)} +\sum_ {n=0}^{2} \Omega^{(n)}.
\end{equation}
Here $b_2^{(1)}$ is the sum of all terms linear in $Y$, and $\Omega^{(n)}$ is the sum of terms with $n$ derivatives which consists of only spin-one fields, entering into $\Gamma^\alpha$. 

For $b_2^{(1)}$ we have (see Eq.\,(\ref{b21final}))
\begin{equation}
\label{om1}
b_2^{(1)}= \mbox{tr}_f \left[ NY-\frac{i}{3}(R\circ \Gamma^\alpha )\! \stackrel{\leftrightarrow}{\partial^\alpha}\! Y\right].
\end{equation}

The terms with two derivatives are
\begin{equation}
\label{om2}
\Omega^{(2)}= \frac{1}{24}  \mbox{tr}_f\! \left[F^{\alpha\beta}\left( S\circ F^{\alpha\beta}\right)
+6(\partial\Gamma )\left( S \circ \partial\Gamma\right)\right].
\end{equation}
The first term makes an additional contribution to the kinetic part of the effective Lagrangian of spin-1 fields described by the second term in ($\ref{b2}$). Recall, that the symmetric matrix $S$ is defined by $S_{ij} = J_{ij} -J_{iijj}$ (\ref{DJ1}). The shorthand $\partial\Gamma\equiv\partial^\alpha\Gamma^\alpha$  implies summation over omitted indices $\alpha$. In applications, one can omit the second term in (\ref{om2}) to ensure that the energy of the massive spin-1 field is positive definite. 

The terms with one derivative can be collected in the expression
\begin{eqnarray}
\label{om1der}
&&\Omega^{(1)} = \frac{i}{3}\mbox{tr}_f\!\left[ H^{\alpha\beta} F^{\alpha\beta}+\frac{1}{2}\,\delta_{\alpha\beta\gamma\sigma}(K\circ \partial^\alpha\Gamma^\beta )E^{\gamma\sigma} \right] \nonumber \\  
&&+ \frac{i}{12}\sum_{i\neq j} R_{ij} \left[8(\partial\Gamma_{ij})(\Gamma^2)_{ji}  
+(F^{\alpha\beta}_{ji} + 9\partial^\alpha\Gamma^\beta_{ji}) L^{\beta\alpha}_{+ij}    \right.   \nonumber \\
&&\left. +\Gamma^\alpha_{ji}\Gamma^\beta_{ij}F^{\alpha\beta}_{jj} \right] + \frac{i}{8}\sum_{i\neq j} S_{ij} (\Gamma^\alpha_{ii}-\Gamma^\alpha_{jj})  \nonumber \\
&&\times \left[ (\Gamma^\alpha_{ij}\! \stackrel{\leftrightarrow}{\partial^\beta}\! \Gamma^\beta_{ji})-(\Gamma^\alpha_{ji}\! \stackrel{\leftrightarrow}{\partial^\beta}\! \Gamma^\beta_{ij} ) + (\Gamma^\beta_{ij}\! \stackrel{\leftrightarrow}{\partial^\alpha}\! \Gamma^\beta_{ji} ) \right] ,  
\end{eqnarray}
which represents the effective three-particle vertices describing the local interaction of vector and axial-vector fields. 

The term without derivatives $\Omega^{(0)}$, which is a sum of the terms $\propto\Gamma^4$, we represent in the form
\begin{equation}
\Omega^{(0)}=\Omega^{(0)}_1+\Omega^{(0)}_2,
\end{equation}
where $\Omega^{(0)}_1$ is given by Eq.\,(\ref{G33}), and 
\begin{eqnarray}
&&\Omega^{(0)}_2=\!\! \sum_{i\neq j\neq k}\!\!
R_{ij} \left[ \left(L^{\alpha\alpha}_- \right)_{ji} \left(\Gamma_{ik}\Gamma_{kj}\right)
         +\left(\Gamma_{ji}\Gamma_{ij}\right)\left(\Gamma_{jk}\Gamma_{kj}\right)
          \right] \nonumber \\
&&+ \frac{1}{3} \delta_{\alpha\beta\gamma\sigma} 
\sum_{i\neq j}  S_{ij}
\left(\Gamma^\alpha_{ii} \Gamma^\beta_{ii} \Gamma^\gamma_{ij} \Gamma^\sigma_{ji} - 
\Gamma^\alpha_{ii} \Gamma^\beta_{ij} \Gamma^\gamma_{jj} \Gamma^\sigma_{ji} \right. \nonumber \\
&&\ \ \ \ \ \ \ \ \ \ \ \ \ \ \ \ \ \ \ \ \left. -\frac{1}{2} \Gamma^\alpha_{ij} \Gamma^\beta_{ji} \Gamma^\gamma_{ij} \Gamma^\sigma_{ji}\right)   \,.
\end{eqnarray}
Here, the expression in the parentheses $\left(\Gamma_{ij}\Gamma_{ji}\right)$ is understood to be summed over alpha $\Gamma^\alpha_{ij}\Gamma^\alpha_{ji}$. 

\section{Continuation to Minkowski space} 
\label{C-M-s}

Let us write down the Lagrangian of the effective theory arising from the considered $1/M$ expansion (\ref{WE}) after continuation to Minkowski space-time
\begin{equation}
\label{effLag}
\mathcal L_{\mbox{\footnotesize eff}}=-\frac{N_c}{32\pi^2} \sum_{n=1}^\infty \mbox{tr}_D\, b_n(x,x)|_M. 
\end{equation}
The index $M$ explicitly indicates that coefficients $b_n$ belong to Minkowski space. The rules of such continuation are given in Appendix \ref{app1}.

The coefficient $b_1(x,x)$ given by Eq.\,(\ref{b1final}) in Minkowski space-time has the following form
\begin{equation}
\label{b1Mink}
b_1|_M= \mbox{tr}_f\left[ J_0\circ (-Y) - \frac{1}{4} (\Delta J_0 \circ \Gamma^\mu ) \Gamma_\mu\right].
\end{equation}
 In particular, $\Gamma^\mu=v^\mu + \gamma^5 a^\mu$, and
\begin{eqnarray}
\label{YMink}
Y&=& s^2 -\{s, M\}+p^2+i\gamma^{5} [s-M, p] \nonumber \\
&-&\frac{i}{4}[\gamma^\mu, \gamma^\nu ] (v_{\mu\nu}+\gamma^{5}a_{\mu\nu})\nonumber \\
&-&i\gamma^\mu (\nabla_{\!\mu} s +i \gamma^{5} \nabla_{\!\mu} p)\,,
\end{eqnarray}
where
\begin{eqnarray}
v_{\mu\nu}&=&\partial_\mu v_\nu -\partial_\nu v_\mu -i [v_\mu, v_\nu ] -i [a_\mu , a_\nu ]\,, \label{eq:b:1Mink}  \\ 
a_{\mu\nu}&=&\partial_\mu a_\nu -\partial_\nu a_\mu -i [v_\mu, a_\nu ] -i [a_\mu, v_\nu ]\,, \label{eq:b:2Mink}  \\ 
\nabla_{\!\mu} s& =&\partial_\mu s -i[v_\mu , s-M]- \{a_\mu, p\}\,, \label{eq:b:3Mink}  \\ 
\nabla_{\!\mu} p&=&\partial_\mu p -i[v_\mu , p]+\{a_\mu , s-M\} \label{eq:b:4Mink}\,. 
\end{eqnarray}

The coefficient $b_2(x,x)$ in Minkowski space-time can be written as 
\begin{equation}
\label{b2Mink}
b_2|_M=\frac{1}{2} \mbox{tr}_f\! \left[ Y\left(J\circ Y\right)-\frac{1}{6} \Gamma^{\mu\nu} \left(J\circ \Gamma_{\mu\nu} \right)\right]+\Delta b_2,
\end{equation}
where 
\begin{equation}
 \label{GMuNu}   
 \Gamma^{\mu\nu}=\partial^\mu\Gamma^\nu -\partial^\nu\Gamma^\mu - i[\Gamma^\mu , \Gamma^\nu ]
 \end{equation}  
is a representation of (\ref{GAlBe}) in Minkowski space-time. For the contributions to $\Delta b_2$ we have
\begin{equation}
\label{b2(1)Mink}
b_2^{(1)}|_M= \mbox{tr}_f \left[ NY-\frac{i}{3}(R\circ \Gamma^\mu )\! \stackrel{\leftrightarrow}{\partial_\mu}\! Y\right].
\end{equation}

In a similar way we find
\begin{equation}
\label{om2Mink}
\Omega^{(2)}_{|_M}= \frac{1}{24}  \mbox{tr}_f\! \left[F^{\mu\nu}\left( S\circ F_{\mu\nu}\right)
+6(\partial\Gamma )\left( S \circ \partial\Gamma\right)\right],
\end{equation} 
where $(\partial\Gamma )$ is understood now as $\partial_\mu\Gamma^\mu$, and
\begin{eqnarray}
\label{om1Mink}
&&\Omega^{(1)}_{|_M} = -\frac{i}{3}\mbox{tr}_f\!\left[ H_{\mu\nu} F^{\mu\nu}+\frac{1}{2}\, g_{\mu\nu\tilde\mu\tilde\nu}(K\circ \partial^\mu\Gamma^\nu )E^{\tilde\mu\tilde\nu} \right] \nonumber \\  
&&+ \frac{i}{12}\sum_{i\neq j} R_{ij} \left[8(\partial\Gamma_{ij})(\Gamma^2)_{ji}  
-(F^{\mu\nu}_{ji} + 9\partial^\mu\Gamma^\nu_{ji}) L^{\nu\mu}_{+ij}    \right.   \nonumber \\
&&\left. -\Gamma^\mu_{ji}\Gamma^\nu_{ij}F^{\mu\nu}_{jj} \right] - \frac{i}{8}\sum_{i\neq j} S_{ij} (\Gamma^\mu_{ii}-\Gamma^\mu_{jj})  \nonumber \\
&&\times \left[ (\Gamma^\mu_{ij}\! \stackrel{\leftrightarrow}{\partial^\nu}\! \Gamma^\nu_{ji})-(\Gamma^\mu_{ji}\! \stackrel{\leftrightarrow}{\partial^\nu}\! \Gamma^\nu_{ij} ) + (\Gamma^\nu_{ij}\! \stackrel{\leftrightarrow}{\partial^\mu}\! \Gamma^\nu_{ji} ) \right], 
\end{eqnarray}
where $g_{\mu\nu\tilde\mu\tilde\nu}=g_{\mu\nu}g_{\tilde\mu\tilde\nu}+g_{\mu\tilde\mu}g_{\nu\tilde\nu}+g_{\mu\tilde\nu}g_{\nu\tilde\mu}$. We will not write out here an expression for $\Omega^{(0)}$, which contains the terms $\propto\Gamma^4$, since here, to go to the Minkowski space, it is sufficient to replace the indices, while the sign in front of any term remains unchanged.

\section{Conclusions} 
\label{Conclusions}
Above, a new method for obtaining the effective low-energy Lagrangian in a theory with heavy particles of unequal masses and a non-trivial flavor symmetry group has been presented in a full detail. Unlike \cite{Min:82}, where authors use the modified DeWitt WKB form and solve recurrent Schr\"odinger-like equations to determine the heat coefficients, our calculations are based on the formula (\ref{alg}), which allows us to resum the proper-time series in accordance with the contributions of corresponding one-loop Feynman diagrams with the required number of external fields. As a result, we arrive at an effective action (\ref{WE}), in which each of the heat coefficients can be calculated independently, and corresponds to a certain order in $1/M^2$ expansion. We have explicitly calculated two leading coefficients $b_1(x,x)$ and $b_2(x,x)$ for the case of the broken $U(3)\times U(3)$ chiral symmetry. The result can be used to study the consequences of $SU(3)$ symmetry violations, for instance, in the framework of the NJL-type models, where, until now, rather rough approximations are used to describe this important effect. Although we have limited ourselves to a particular example, the proposed approach can be easily extended to the case of an arbitrary internal symmetry group, for example, to study an explicit and spontaneous symmetry violation within the context of effective field theories beyond the Standard Model.

\section*{Acknowledgments}
I am grateful to Prof. B. Hiller for the interest in the subject and useful discussions. This work is supported by Grant from Funda\c{c}\~ ao para a Ci\^ encia e Tecnologia (FCT) through the Grant No. CERN/FIS-COM/0035/2019, and the European Cooperation in Science and Technology organisation through the COST Action CA16201 program.

\appendix
\section{The Wick rotation to the Euclidean space-time}
\label{app1}
In this Appendix,  we collect some useful formulas and conventions made to perform a Wick rotation of the time axis to transit from Minkowski space-time with the metric tensor $g_{\mu\nu}=\mbox{diag}\, (1,-1,-1,-1)$  to Euclidean space with metric $\delta_{\alpha\beta}=\mbox{diag}\, (1,1,1,1)$ and back. Here the first two letters of Greek alphabet $\alpha, \beta$ are used to enumerate the coordinates in  Euclidean space $\alpha,\beta =1,2,3,4$, on the contrary, the letters $\mu,\nu$, run over the set $\mu,\nu =0,1,2,3$ in Minkowski space-time.  

For coordinates $x^\mu$ and the Lorentz four-vector $v^\mu$ we apply the following correspondence rules with the Euclidean coordinates $x_\alpha$ and the vector components $v_\alpha$
\begin{eqnarray}
x^\mu &=&(x^0,\vec x)\to x_\alpha =(\vec x,x_4)\,,  \!\!\quad x^0\to -ix_4\,. \label{eq:a:1} \\
\partial_\mu &=&(\partial_0, \vec\nabla )\to \partial_\alpha =(\vec\nabla ,\partial_4)\,, \!\!
\quad \ \partial_0\to i\partial_4\,. \label{eq:a:2}\\
v^\mu &=&(v^0,\vec v)\to v_\alpha =(\vec v, v_4)\,, \quad v^0\to -iv_4\,. \label{eq:a:3}
\end{eqnarray}
It follows then from (\ref{eq:a:2}) and (\ref{eq:a:3}) that
\begin{eqnarray}
&&v^\mu v_\mu =v_0^2-\vec v^{\,2}\to -v_4^2-\vec v^{\,2}=-v_\alpha^2\,, \label{eq:a:4}  \\ 
&&v^\mu\partial_\mu=v_0\partial_0+\vec v\,\vec\nabla\to v_\alpha\partial_\alpha\,. \label{eq:a:5}  
\end{eqnarray} 

For the Dirac $\gamma$-matrices we use the following conventions
\begin{equation}
\gamma^\mu =(\gamma^0, \vec\gamma )\to \gamma_\alpha=(\vec\gamma,\gamma_4)\,, \quad 
\gamma^0\to -i\gamma_4\,.  
\end{equation}
Our choice corresponds to the case when $O(4)$ gamma matrices $\gamma_\alpha$ in the four-dimensional Euclidean space are anti-Hermitian. Indeed, from the standard property $\gamma^{\mu\dagger}=(\gamma^0, -\vec\gamma)$ and definitions made above one finds that $\gamma_\alpha^\dagger=-\gamma_\alpha$. As a consequence, the Euclidean Clifford algebra emerging from $\{\gamma^\mu,\gamma^\nu\}=2g^{\mu\nu}$ necessitates a change in the defining equation $\{\gamma_\alpha,\gamma_\beta\}=-2\delta_{\alpha\beta}$. We have also that $\gamma^\mu\partial_\mu\to\gamma_\alpha \partial_\alpha$, $\gamma^\mu v_\mu\to -\gamma_\alpha v_\alpha$. 

The other consequence is that $\gamma^5=-i\gamma^0\gamma^1\gamma^2\gamma^3$ matrix can be defined in Euclidean space as $\gamma_{5E}=\gamma_1\gamma_2\gamma_3\gamma_4$. It follows then that $\gamma^5\to \gamma_{5E}$, and both matrices are Hermitian. 

For the Dirac operator 
\begin{equation}
D=i\gamma^\mu d_\mu -M + s+i\gamma^5 p, 
\end{equation}
where $d_\mu =\partial_\mu -i\Gamma_\mu$, and $\Gamma_\mu = v_\mu +\gamma^5 a_\mu$, one finds 
\begin{equation}
D\to D_E^{}=i\gamma_\alpha d_\alpha-M+s+i\gamma_{5E} p,
\end{equation}
where $d_\alpha=\partial_\alpha +i\Gamma_\alpha\,$, $\Gamma_\alpha =v_\alpha +\gamma_{5E}a_\alpha$. 

This yields 
\begin{eqnarray} 
\label{D2}
&&D_E^\dagger D_E^{}= -d_\alpha^2 +(s-M)^2 +p^2 +i\gamma_{5E} [s-M, p] \nonumber \\
&&\ \ \ \   +\frac{1}{4}[\gamma_\alpha, \gamma_\beta] [d_\alpha, d_\beta] 
+ i[\gamma_\alpha (s+i\gamma_{5E} p -M), d_\alpha] \nonumber \\
&&\ \ \ \ =M^2-d^2 +Y\,,
\end{eqnarray}
where $d^2=d_\alpha d_\alpha$, 
\begin{eqnarray}
\label{Y}
Y&=& s^2 -\{s, M\}+p^2+i\gamma_{5E} [s-M, p] \nonumber \\
&+&\frac{i}{4}[\gamma_\alpha, \gamma_\beta] (v_{\alpha\beta}+\gamma_{5E}a_{\alpha\beta})\nonumber \\
&-&i\gamma_\alpha (\nabla_{\!\alpha} s +i \gamma_{5E} \nabla_{\!\alpha} p)\,,
\end{eqnarray}
and the chiral-covariant objects are
\begin{eqnarray}
v_{\alpha\beta}&=&\partial_\alpha v_\beta -\partial_\beta v_\alpha +i [v_\alpha, v_\beta] +i [a_\alpha, a_\beta]\,, \label{eq:b:1}  \\ 
a_{\alpha\beta}&=&\partial_\alpha a_\beta -\partial_\beta a_\alpha +i [v_\alpha, a_\beta] +i [a_\alpha, v_\beta]\,, \label{eq:b:2}  \\ 
\nabla_{\!\alpha} s& =&\partial_\alpha s +i[v_\alpha, s-M]+\{a_\alpha, p\}\,, \label{eq:b:3}  \\ 
\nabla_{\!\alpha} p&=&\partial_\alpha p +i[v_\alpha, p]-\{a_\alpha, s-M\} \label{eq:b:4}\,. 
\end{eqnarray}
We have given these well-known formulas here both for the completeness of our presentation and to clarify notations used in the main text of our work.

\section{The modulus of the determinant}
\label{app2}

This is a summary of the standard steps to evaluate the modulus of determinant (\ref{logdet}). The self-adjoint elliptic operator $\Delta =D_E^\dagger D_E^{}$ can be treated as an operator acting on a fictitious Hilbert space, where the quantum mechanical algebra of coordinates and momenta is realized, i.e., where $|x\rangle$ is an eigenvector of a commuting set of Hermitian operators $\hat{x}_\alpha$ such that $\hat{x}_\alpha |x\rangle =x_\alpha |x\rangle$ and $\langle x|y\rangle =\delta (x-y)$. The Hermitian operators, $\hat{p}_\alpha =-i\partial_\alpha$,  which are conjugate to $\hat{x}_\alpha$, obey canonical commutation relations: $[\hat{x}_\alpha ,\hat{p}_\beta ]=i\delta_{\alpha\beta},\ [\hat{x}_\alpha ,\hat{x}_\beta ]=[\hat{p}_\alpha ,\hat{p}_\beta]=0$. 

Let us consider an eigenket $|p\rangle$ of $\hat{p}_\alpha$. Its representative in the Schr\"{o}dinger picture is $\langle x|p\rangle =(2\pi )^{-2}\exp (ix_\alpha p_\alpha )$. Using this plane wave basis one can describe the action of the operator $e^{-t\Delta}$ on the Hilbert space by the heat kernel function of $\Delta$
\begin{equation}
   h(x,y;t)=\langle x|e^{-t\Delta}|y\rangle =\!\int\! d^4p\, \langle x|e^{-t\Delta}|p\rangle\langle p|y\rangle.
\end{equation}

Taking into account the relations: $\langle x|\hat p_\alpha |p\rangle =p_\alpha  \langle x|p\rangle$ $= -i\partial^{x}_\alpha \langle x|p\rangle$ and $\langle x|\hat{x}_\alpha |p\rangle =x_\alpha\langle x|p\rangle$ one can proceed
\begin{eqnarray}
&&\!\!\!\!\!\!\!\!\!\!\!\!\!\! \langle x|e^{-t\Delta}|y\rangle=\!\int\! d^4p\,  e^{-t\Delta_x}\langle x|p\rangle \langle p|y\rangle  \nonumber\\
&&=\!\int\!\frac{d^4p}{(2\pi )^4}e^{-ipy}e^{-t\Delta_x}e^{ipx} \nonumber \\
&&=\!\int\!\frac{d^4p}{(2\pi )^4}e^{ip(x-y)}\left(e^{-ipx}e^{-t\Delta_x}e^{ipx}\right), 
\end{eqnarray}
where $\Delta_x$ is the operator $\Delta$ in the coordinate representation.

Now note that
\begin{eqnarray}
&&e^{-ipx}e^{-t\Delta_x}e^{ipx}=\sum_{n=0}^\infty \frac{(-t)^n}{n!}e^{-ipx}(\Delta_x)^ne^{ipx} \nonumber \\
&&=\sum_{n=0}^\infty \frac{(-t)^n}{n!}\left(e^{-ipx}\Delta_xe^{ipx}\right)^n,
\end{eqnarray}
where 
\begin{eqnarray}
e^{-ipx}\Delta_xe^{ipx}&=&\Delta_x-\left[ipx, \Delta_x\right]+\frac{1}{2!}[ipx,[ipx, \Delta_x]]\nonumber \\
&=&{\Delta}_x-2ip_\alpha d_\alpha +p^2.
\end{eqnarray}
Since the operator $\Delta_x$ contains derivatives of at most second order, its higher order commutators with $ipx$ are equal to zero.  At the last stage we used Eq.(\ref{D2}), namely
\begin{eqnarray*} 
[ipx, \Delta_x]
&=&[ipx, M^2-d^2+Y]=-[ipx, d^2] \\
&=&-[ipx, \partial_\alpha^2 + 2i\Gamma_\alpha \partial_\alpha] \\
&=&-[ipx,\partial_\alpha] \partial_\alpha -\partial_\alpha [ipx, \partial_\alpha] -2i\Gamma_\alpha [ipx,\partial_\alpha] \\
&=& 2ip_\alpha d_\alpha\,, \\
&&\!\!\!\!\!\!\!\!\!\!\!\!\!\!\!\!\!\!\!\!\!\!\!\!\!\!\!\!\!\!\! [ipx, [ipx, \Delta_x]]=[ipx, 2ip_\alpha d_\alpha]=2p_\alpha^2\,.
\end{eqnarray*}

As a result we obtain 
\begin{eqnarray}
&&\langle x|e^{-t\Delta}|y\rangle =\int\!\!\frac{d^4p}{(2\pi )^4}e^{ip(x-y)}e^{-tp^2}e^{-t({\Delta}_x-2ipd)} 
\nonumber \\   
&&\quad \!\!=\!\!\int\!\!\frac{d^4p}{(4\pi^2t)^2}e^{-p^2+\frac{i}{\sqrt{t}}p(x-y)} e^{-t\left(\Delta_x-\frac{2ipd}{\sqrt{t}}\right)},
\end{eqnarray}
or, for the trace
\begin{eqnarray}
\label{hk}
\mbox{Tr}(e^{-t\Delta})&=&\!\int\! d^4x\,\mbox{tr}_I\langle x|e^{-t\Delta}|x\rangle \nonumber \\ 
&=&\int\!\frac{d^4xd^4p}{(4\pi^2t)^2}\,e^{-p^2}\,\mbox{tr}_I\, e^{-t\left(\Delta_x-\frac{2ipd}{\sqrt{t}}\right)},
\end{eqnarray}
where tr$_I$ means the trace over color, isospin, and Dirac indices. Expanding the last exponent in the series in the proper-time, one can take the integrals over four-momenta with the help of the formulas 
\begin{equation}
\int\!\! \frac{d^4p}{(2\pi)^4}\,e^{-p^2} p_{\alpha_1} p_{\alpha_2}\ldots p_{\alpha_{2n}}=\frac{\delta_{\alpha_1\alpha_2\ldots\alpha_{2n}}}{2^n(4\pi)^2}\,.
\end{equation} 
Note that only even number of internal momenta $p_\alpha$ contribute, the odd number, obviously, gives zero; the totally symmetric tensor $\delta_{\alpha_1\alpha_2\ldots\alpha_{2n}}$ is a product of Kronecker's deltas $\delta_{\alpha_1\alpha_2}\delta_{\alpha_3\alpha_4}\ldots\delta_{\alpha_{2n-1}\alpha_{2n}}$ plus all possible permutations of indices; there are $(2n-1)!!$ terms in total. For example, 
\begin{equation}
\label{daaaa}
\delta_{\alpha_1\alpha_2\alpha_3\alpha_4}=\delta_{\alpha_1}^{\alpha_2}\delta^{\alpha_3}_{\alpha_4}+\delta_{\alpha_1}^{\alpha_3}\delta^{\alpha_2}_{\alpha_4}+\delta_{\alpha_1}^{\alpha_4}\delta_{\alpha_2}^{\alpha_3}.
\end{equation} 
Finally, combining (\ref{logdet}) and (\ref{hk}) we come to the Eq.\,(\ref{logdet2}).

\section{Proof of the formula (\ref{c1n})}
\label{app3}
Here we present some details related with obtaining of Eq.\,(\ref{c1n}). Let us first consider the case when $n=1$
\begin{eqnarray}
&&\mbox{tr}_f \left[e^{-tM^2} f_1(t,A)\right]\nonumber \\
&&=\sum_{i,j,k}^3 e^{-tm_i^2} \int_0^t ds \, e^{s(m_j^2-m_k^2)} \mbox{tr}_f (E_iE_jAE_k) \nonumber \\
&&=\sum_i t e^{-tm^2_i}  \mbox{tr}_f A_i.
\end{eqnarray}
Hence, 
\begin{equation}
c_i(t)=e^{-tm_i^2}.
\end{equation}

For $n=2$ we have 
\begin{eqnarray}
&&\mbox{tr}_f \left[e^{-tM^2} f_2(t,A)\right]=\sum_{i,j}^3 \int_0^t ds_1 \int_0^{s_1} ds_2      \nonumber \\
&& \times e^{(s_1-s_2-t)m_i^2} e^{(s_2-s_1)m_j^2} \mbox{tr}_f (A_iA_j) \nonumber\\
&&=\sum_{i,j}^3\left(t\,\frac{e^{-tm^2_i}}{\Delta_{ji}} + \frac{e^{-tm^2_j}-e^{-tm_i^2}}{\Delta_{ij}^2} \right)
\mbox{tr}_f (A_iA_j) \nonumber \\
&&= \sum_{i,j}^3 \frac{t}{2\Delta_{ji}}  \left(e^{-tm_i^2}-e^{-tm_j^2}\right) \mbox{tr}_f (A_iA_j).
\end{eqnarray}
Here $\Delta_{ij}=m_i^2-m_j^2$, and on the last stage we used that trace is symmetric under an exchange $i\leftrightarrow j$. It gives 
\begin{equation}
c_{i_1i_2}(t)=\frac{1}{t}\sum_{perm} \frac{c_{i_1}(t)}{\Delta_{i_2i_1}}\,,
\end{equation}
the sum includes two terms obtained by cyclically permuting the indices $(i_1,i_2)\to (i_2,i_1) $.

For $n=3$ 
\begin{eqnarray}
&&\mbox{tr}_f \left[e^{-tM^2} f_3(t,A)\right]=\sum_{i,j,k}^3 \int_0^t\!\! ds_1\! \int_0^{s_1}\!\!\! ds_2 \!  \int_0^{s_2} \!\!\! ds_3     \nonumber \\
&& \times e^{(s_1-s_3-t)m_i^2} e^{(s_2-s_1)m_j^2} e^{(s_3-s_2)m_k^2} \mbox{tr}_f (A_iA_jA_k) \nonumber \\
&&=\sum_{i,j,k}^3 \left(\frac{tc_i(t)}{\Delta_{ij}\Delta_{ik}}+ \frac{c_i(t)-c_j(t)}{\Delta_{ij}^2\Delta_{jk}}+  \frac{c_i(t)-c_k(t)}{\Delta_{ik}^2\Delta_{kj}}  \right) \nonumber\\
&&\times\, \mbox{tr}_f (A_iA_jA_k).
\end{eqnarray}
Since the trace is symmetric with respect to the cyclic permutation of matrices, one can add to this expression two others, obtained by cyclic permutations of indices $(i,j,k)\to (j,k,i)$ and $(i,j,k)\to (k,i,j)$, and the result should be divided by three. As a result, only the first term in parentheses will lead to the non zero contribution. This yields the following expression for the coefficient   
\begin{equation}
c_{i_1i_2i_3}(t)=\frac{2!}{t^2}\sum_{perm} \frac{c_{i_1}(t)}{\Delta_{i_2i_1}\Delta_{i_3i_1}}\, .    
\end{equation}

Consider the case $n=4$. Here we have
\begin{eqnarray}
&&\mbox{tr}_f \left[e^{-tM^2} f_4(t,A)\right]=\sum_{i,j,k,l}^3 \int_0^t\!\! ds_1\! \int_0^{s_1}\!\!\! ds_2 \!  \int_0^{s_2} \!\!\! ds_3  \! \int_0^{s_3}\!\!\! ds_4    \nonumber \\
&&\times e^{(s_1-s_4-t)m_i^2} e^{(s_2-s_1)m_j^2} e^{(s_3-s_2)m_k^2} e^{(s_4-s_3)m_l^2} \nonumber \\
&&\times \mbox{tr}_f (A_iA_jA_kA_l) \nonumber \\
&&=\sum_{i,j,k,l}^3  \left( \frac{tc_i(t)}{\Delta_{ji}\Delta_{ki}\Delta_{li}} + \frac{c_j(t)-c_i(t)}{\Delta_{ij}^2\Delta_{kj}\Delta_{lj}}   + \frac{c_k(t)-c_i(t)}{\Delta_{ik}^2\Delta_{lk}\Delta_{jk}}  \right.\nonumber \\
&&\left.  + \frac{c_l(t)-c_i(t)}{\Delta_{il}^2\Delta_{kl}\Delta_{jl}} \right) \mbox{tr}_f (A_iA_jA_kA_l).
\end{eqnarray}
By using again the cyclic permutation property of trace, one can easily find out that only the first term in parentheses will contribute with a non zero result. Here we have four cyclic permutations of indices, correspondingly it gives four terms for the coefficient 
\begin{equation}
c_{i_1i_2i_3i_4}(t)=\frac{3!}{t^3} \sum_{perm} \frac{c_{i_1}(t)}{\Delta_{i_2i_1}\Delta_{i_3i_1}\Delta_{i_4i_1}}\,.
\end{equation}

Continuing the above procedure, we can find by induction an arbitrary coefficient in formula (\ref{c1n})
\begin{equation}
c_{i_1i_2\ldots i_n}(t)=\frac{(n-1)!}{t^{n-1}}\!\! \sum_{perm} \!\frac{c_{i_1}(t)}{\Delta_{i_2i_1}\Delta_{i_3i_1}\ldots \Delta_{i_ni_1}}\, .
\end{equation}
Although the sum includes $n$ terms obtained by cyclically permuting the indices $i_1, i_2, \ldots , i_n$, it is easy to show that the coefficients $c_{i_1i_2\ldots i_n}$ are {\it totally} symmetric.

One can also establish a recurrence formula relating coefficients of order $n$, $c_{i_1i_2\ldots i_n}(t)$, for $n>1$, with coefficients of order $n-1$  
\begin{equation}
\label{recrel}
c_{i_1i_2\ldots i_n}(t)=\frac{n-1}{t\Delta_{i_{n-1}i_{n}}}\left[c_{i_1i_2\ldots i_{n-2}i_n}(t)-c_{i_1i_2\ldots i_{n-1}}(t)\right]. 
\end{equation}
This relation follows from the following easily verified identities
\begin{eqnarray}
&&\sum_{perm} \frac{1}{\Delta_{i_2i_1}}=0,   \\
&&\sum_{perm} \frac{1}{\Delta_{i_2i_1}\Delta_{i_3i_1}}=0,    \\
&& \frac{}{} \ldots\ldots\ldots\ldots\ldots\ldots\ldots  {} \nonumber \\
&&\sum_{perm} \!\frac{1}{\Delta_{i_2i_1}\Delta_{i_3i_1}\ldots \Delta_{i_ni_1}}=0. 
\end{eqnarray}

Recurrent formula (\ref{recrel}), inter alia, simplifies the proof of the relations 
\begin{equation}
\label{eqind}
c_i(t) = c_{ii}(t) =c_ {ii\ldots i}(t)\,.
\end{equation}
Indeed, one has, for instance, 
\begin{eqnarray}
c_{ii}(t)&=&\lim_{m_i\to m_j} c_{ij}(t)=\lim_{m_i\to m_j} \frac{e^{-tm_j^2}-e^{-tm_i^2}}{t\Delta_{ij}} \nonumber\\
&=&\lim_{m_i\to m_j} e^{-tm_i^2} \frac{e^{t\Delta_{ij}}-1}{t\Delta_{ij}}=c_i(t).
\end{eqnarray} 
Using (\ref{recrel}) we can extend this to any $c_ {ii\ldots i}(t)$ case.

\section{Integrals over proper-time}
\label{app4}
If the masses of the Fermi fields are non-degenerate, integration over the proper-time leads to expressions depending on the specific masses of the constituent quarks in the corresponding Feynman diagram. Accordingly, the number of different loop integrals increases. We will classify the integrals in accord with the number of internal quark lines presented in the loop  diagram. Since we are dealing with an $1/M$ expansion, the integrals presented below give only the leading terms of this expansion. Along with separating the main divergences of the Feynman diagrams, such integrals contain finite contributions that vanish in the case of equal quark masses. These finite contributions are uniquely determined by the coefficients $c_{i_1i_2\ldots i_n}$ that are completely symmetric with respect to the permutation of any two indices.

\subsection{Tadpole diagrams}  
The diagrams with only one internal quark line (tadpole) are associated with the proper-time integral
\begin{equation}
\int\!\frac{dt}{t^2}\, \rho_{t,\Lambda} c_{i} (t) =J_0(m_i^2)\,. 
\end{equation}
We will use Pauli-Villars regularization corresponding to two subtractions at the mass-scale $\Lambda$
\begin{equation}
 \rho_{t,\Lambda}=1-(1+t\Lambda^2)e^{-t\Lambda^2}.
\end{equation}
In this case, it follows that
\begin{equation}
\label{J0mi}
J_0(m_i^2)=\Lambda^2-m_i^2 \ln\left(1+\frac{\Lambda^2}{m_i^2}\right)\,.
\end{equation} 
This obviously coincides with (\ref{J0}).

\subsection{Self-energy diagrams}   
The quark-loop self-energy graphs are described by the proper-time integrals
\begin{equation}
\int\!\frac{dt}{t^{2-n}}\, \rho_{t,\Lambda} c_{ij} (t) =J_n(m_i^2,m_j^2)\,. 
\end{equation}
Due to the symmetry of the coefficients $c_{ij}=c_{ji}$ the result of integration is symmetric with respect to the replacement $m_i\leftrightarrow m_j$. Indeed, in the case $n=0$, we have  
\begin{eqnarray}
\label{J0ij}
&&\!\!\!\!\!\!\! J_0(m_i^2,m_j^2)=\frac{\Lambda^2}{2} +\frac{\Lambda^4}{2\Delta_{ij}} \ln\frac{\Lambda^2+m_i^2}{\Lambda^2+m_j^2} \nonumber \\
&&\!\!\!\!\!\!\!\!\! -\frac{1}{2\Delta_{ij}}\left[m_i^4\ln\left(1+\frac{\Lambda^2}{m_i^2}\right)-m_j^4\ln\left(1+\frac{\Lambda^2}{m_j^2}\right)\right]\!,
\end{eqnarray}
where $\Delta_{ij}=m_i^2-m_j^2$\,. In the coincidence limit $m_i\to m_j$, $$\lim_{m_i\to m_j} J_0(m_i^2,m_j^2)=J_0(m_i^2)\,.$$
The other integral for $n=1$ is given by the difference of the tadpole contributions 
\begin{eqnarray}
\label{J-ij}
&&J_1(m_i^2,m_j^2)=\frac{1}{\Delta_{ij}}\left[J_0(m_j^2)-J_0(m_i^2)\right]  \\
&&=\frac{1}{\Delta_{ij}}\left[m_i^2 \ln \left(1+\frac{\Lambda^2}{m_i^2}\right)-m_j^2 \ln \left(1+\frac{\Lambda^2}{m_j^2}\right)\right]. \nonumber
\end{eqnarray}
Since we also have expressions for the similar $J_1$-type contributions coming from the triangle and box diagrams, we will collect them together in one formula 
\begin{equation}
\label{integrals}
\left( \begin{array}{c}
J_{ij} \\  J_{ijk} \\  J_{iijk} \\
\end{array}\right)\equiv \!\!\int\limits^\infty_0  \frac{dt}{t}\,\rho_{t,\Lambda}\, \left(
\begin{array}{c} c_{ij} (t) \\ c_{ijk} (t) \\ c_{iijk} (t) \\ \end{array}\right).
\end{equation}
From Eq.\,(\ref{eqind}), it follows that 
\begin{equation}
J_{ii}=J_{iii}=J_{iiii}=J_1(m_i^2)\equiv J_i.
\end{equation} 
The integrals take this form when all quark masses in the loop are equal.

\subsection{Triangle diagrams}  
The proper-time integral 
\begin{equation}
\int\!\frac{dt}{t}\, \rho_{t,\Lambda} c_{ijk} (t) =J_1(m_i^2,m_j^2,m_k^2)\equiv J_{ijk} 
\end{equation}
describes a local part of the diagram with three internal quark lines. As before, the full symmetry of the coefficient $c_{ijk}$ allows one to conclude that this integral does not change when any two fermion masses are exchanged. To verify this, we use the formula (\ref{recrel}), which at the first stage of calculations leads to a seemingly asymmetric result
\begin{equation}
J_{ijk} =\frac{2}{\Delta_{jk}}\left[J_0(m_i^2,m_k^2)-J_0(m_i^2,m_j^2) \right],
\end{equation} 
where $J_0(m_i^2,m_j^2)$ is given by Eq. (\ref{J0ij}). However, the direct calculation of the right hand side of this expression shows explicitly its total symmetry 

\begin{widetext} 
\begin{eqnarray}
&&J_1(m_i^2,m_j^2,m_k^2)=\frac{1}{\Delta_{ij}\Delta_{jk}\Delta_{ki}}\left[\Lambda^4
\left(m_i^2 \ln\frac{\Lambda^2+m_k^2}{\Lambda^2+m_j^2}  +m_j^2 \ln\frac{\Lambda^2+m_i^2}{\Lambda^2+m_k^2} +m_k^2 \ln\frac{\Lambda^2+m_j^2}{\Lambda^2+m_i^2}\right) \right. \nonumber\\ 
&&\left.+m_i^4\Delta_{kj}\ln\left(1+\frac{\Lambda^2}{m_i^2}\right) +m_j^4\Delta_{ik}\ln\left(1+\frac{\Lambda^2}{m_j^2}\right)+m_k^4\Delta_{ji}\ln\left(1+\frac{\Lambda^2}{m_k^2}\right)\right]\,.
\end{eqnarray}
This expression allows to obtain a particular value corresponding to the cases with two equal masses:
\begin{eqnarray}
\label{c9}
J_1(m_i^2,m_i^2,m_j^2)&=& \frac{2}{\Delta_{ij}}\left[J_0(m_i^2,m_j^2)-J_0(m_i^2) \right]=
\frac{1}{\Delta_{ij}}\left[2m_i^2\ln\left(1+\frac{\Lambda^2}{m_i^2}\right)-\Lambda^2\right] \nonumber \\
&+&\frac{1}{\Delta_{ij}^2} 
\left[ m_j^4\ln\left(1+\frac{\Lambda^2}{m_j^2}\right)-m_i^4 \ln\left(1+\frac{\Lambda^2}{m_i^2}\right)+ \Lambda^4 \ln \frac{\Lambda^2+m_i^2}{\Lambda^2+m_j^2} \right]\,.
\end{eqnarray}
\end{widetext}    
It should be noted that in the theory considered, there are only three different flavor states of quarks: $u,d$ and $s$. Therefore, Eq. (\ref{c9}) describes the six possible cases: $uud$, $uus$, $ddu$, $dds$, $ssu$, and $ssd$. Notice also that a contribution, for instance, of $uus$ triangle differs from $ssu$ one. The difference is antisymmetric with respect to the replacement $m_i\leftrightarrow m_j$ 
\begin{equation}
\label{Dj0}
J_{iij} -J_{jji}=\frac{2}{\Delta_{ij}} \Delta J_0(m_i^2,m_j^2),
\end{equation}
where 
\begin{equation}
\label{DJ0}
\Delta J_0(m_i^2,m_j^2) = 2J_0(m_i^2,m_j^2) - J_0(m_i^2) -J_0(m_j^2) \,.
\end{equation}
This value is finite at $\Lambda\to\infty$. The sum is 
\begin{equation}
\label{J1+}
J_{iij} + J_{jji} =2 J_1(m_i^2,m_j^2)\,.
\end{equation}
As a consequence, we have
\begin{equation}
\label{D15}
J_1(m_i^2,m_i^2,m_j^2)=J_1(m_i^2,m_j^2)+ R(m_i^2,m_j^2),
\end{equation}
where antisymmetric part of this expression is defined as
\begin{equation}
\label{Rij}
R(m_i^2,m_j^2) =\frac{\Delta J_0(m_i^2,m_j^2)}{\Delta_{ij}}\equiv R_{ij}\,.
\end{equation}

\subsection{Box diagrams}        
Consider the last of the divergent-type expressions that arise when one calculates the quark box diagrams with external meson fields. These are given by the integrals of the form
\begin{equation} 
\int\!\frac{dt}{t}\, \rho_{t,\Lambda}\, c_{ijkl} (t) =J_1(m_i^2,m_j^2,m_k^2,m_l^2)\,. 
\end{equation}
Since quarks have only three flavors, at least two indices in $c_{ijkl}$ coincide. Therefore, we actually need to calculate the master integral
\begin{equation}
\int\!\frac{dt}{t}\, \rho_{t,\Lambda} c_{iijk} (t) = J_{iijk}\,. 
\end{equation}
To this end, we again use the formula (\ref{recrel}), which in this case takes the form
\begin{equation}
\label{dif}
c_{iijk}(t)=\frac{3}{t\Delta_{kj}}\left[c_{iij}(t)-c_{iik}(t)\right].
\end{equation}
The quadratic divergences arising when one integrates each of the coefficients $c_{iij}(t)$ are canceled in the difference (\ref{dif}). As a result, we arrive at a logarithmically divergent (at $\Lambda\to \infty$) expression 
\begin{widetext}
\begin{eqnarray}
J_{iijk}&=&\frac{1}{\Delta_{ij}\Delta_{ik}}\left[3m_i^4\ln\left(1+\frac{\Lambda^2}{m_i^2}\right)-\Lambda^2(2\Lambda^2+m_i^2)\right] \nonumber \\
&+&\frac{1}{\Delta_{kj}} \left\{ \frac{1}{\Delta_{ij}^2} \left[m_i^6 \ln\left(1+\frac{\Lambda^2}{m_i^2}\right) - m_j^6 \ln\left(1+\frac{\Lambda^2}{m_j^2}\right)+\Lambda^4\left(2\Lambda^2+3m_j^2\right)\ln\frac{\Lambda^2+m_j^2}{\Lambda^2+m_i^2}\right] \right. \nonumber \\
&-&\left.\frac{1}{\Delta_{ik}^2} \left[m_i^6 \ln\left(1+\frac{\Lambda^2}{m_i^2}\right) - m_k^6 \ln\left(1+\frac{\Lambda^2}{m_k^2}\right)+\Lambda^4\left(2\Lambda^2+3m_k^2\right)\ln\frac{\Lambda^2+m_k^2}{\Lambda^2+m_i^2}\right] \right\}\,. 
\end{eqnarray}
Notice, that in full accord with (\ref{dif}) we have   
\begin{equation}
\label{symJiijk}
J_{iijk}=J_{iikj}.
\end{equation}

Cases with two different types of flavors can be obtained from this expression by passing to the corresponding limits of equal masses. There are two of them: $J_{iijj}$ and $J_{iiij}$. As a result, we have
\begin{eqnarray}
\label{J-iijj}
J_{iijj}&=&\frac{1}{\Delta_{ij}^2}\left\{
m_i^4\left(1-2\frac{m_j^2}{\Delta_{ij}}\right) \ln\left(1+\frac{\Lambda^2}{m_i^2}\right)+ m_j^4\left(1-2\frac{m_i^2}{\Delta_{ji}}\right) \ln\left(1+\frac{\Lambda^2}{m_j^2}\right)\right. \nonumber \\
&-&\Lambda^2(4\Lambda^2+m_i^2+m_j^2) -\left.\Lambda^4\left[3+\frac{2}{\Delta_{ij}}(2\Lambda^2+3m_j^2)\right]\ln\frac{\Lambda^2+m_j^2}{\Lambda^2+m_i^2}\right\}\,.
\end{eqnarray}
\begin{eqnarray}
\label{iiij}
J_{iiij}&=&\frac{1}{\Delta_{ij}^3}\left\{
m_i^6 \ln\left(1+\frac{\Lambda^2}{m_i^2}\right) - m_j^6 \ln\left(1+\frac{\Lambda^2}{m_j^2}\right)
+\Delta_{ij}\Lambda^2(2\Lambda^2-m_i^2+2m_j^2)\right. \nonumber \\
& -&3\Delta_{ij}m_i^2m_j^2  \ln\left(1+\frac{\Lambda^2}{m_i^2}\right)
+\left.\Lambda^4( 2\Lambda^2+3m_j^2)\ln\frac{\Lambda^2+m_j^2}{\Lambda^2+m_i^2}\right\}\,.
\end{eqnarray}
\end{widetext}
From (\ref{J-ij}) and (\ref{J-iijj}) we find the relation 
\begin{equation}
J_{iijj} =J_{ij}- S(m_i^2,m_j^2), 
\end{equation}
where
\begin{eqnarray}
\label{DJ1}
S(m_i^2,m_j^2)&=& \frac{1}{\Delta_{ij}^3}\left\{ m_i^2m_j^2 (m_i^2+m_j^2)\ln\frac{(\Lambda^2+m_i^2)m_j^2}{(\Lambda^2+m_j^2)m_i^2} \right. \nonumber \\
&+&\left.\Lambda^4\left[4\Lambda^2 + 3(m_i^2+m_j^2)\right]\ln\frac{\Lambda^2+m_j^2}{\Lambda^2+m_i^2}\right\}  \nonumber \\
&+&\frac{\Lambda^2}{\Delta_{ij}^2} (4\Lambda^2 +m_i^2+m_j^2).   
\end{eqnarray}
The latter expression is symmetric under exchange of masses and finite at $\Lambda\to \infty$.

Eq.\,(\ref{iiij}) describes the six possible variants of the distribution of quarks along the inner lines of the box diagram, these are $uuud$, $uuus$, $dddu$, $ddds$, $sssu$, and $sssd$. In particular, it follows that
\begin{equation}
\label{iiij-}
J_{iiij}-J_{jjji}=3R_{ij}\,,
\end{equation}
where $R_{ij}$ is given by (\ref{Rij}). On the other hand, the sum of these integrals is
\begin{equation}
\label{iiij+}
J_{iiij}+J_{jjji}=2J_1(m_i^2,m_j^2)+S(m_i^2,m_j^2)\,.
\end{equation}

\section{Mixed sums for $b_n(x,x)$}
\label{app5}
In the main text we make use summations which contain both the mass-dependent remnant, $J_{i_1 i_2 \ldots i_n}$, of one-loop quark integrals and the corresponding field-dependent coefficients $T_{i_1 i_2 \ldots i_n}$. The result of such summations depends on the properties of $J_{i_1 i_2 \ldots i_n}$ which were established in Appendix \ref{app4}. Here some important results on these sums are presented. 

The simplest case is given by the sum 
\begin{eqnarray}
\label{simplestsum}
\sum_{i,j} J_{ij}T_{ij}&=&\sum_{i,j} J_{ij}T_{(ij)} \nonumber \\
&=&\sum_{i=1}^3 J_{i}T_{ii}+2 \sum_{i<j} J_{ij}T_{(ij)},
\end{eqnarray}
where 
\begin{equation}
T_{(ij)}=\frac{1}{2}\left(T_{ij}+T_{ji}\right).
\end{equation}

Consider now the sum related with the triangle diagrams. The corresponding one-loop integral induces at leading order of its large mass expansion the function $J_{ijk}$, which is known to be completely symmetric under any exchange of masses. Due to this symmetry we have     
\begin{eqnarray}
\label{trdiag}
&&\sum_{i,j,k} J_{ijk} \, T_{ijk} = \sum_{i,j,k} J_{ijk} \, T_{(ijk)} \\
&&=\sum_i J_{i}T_{iii}+3\sum_{i\neq j} J_{iij}T_{(iij)} \nonumber 
+6 J_{123} T_{(123)},
\end{eqnarray}
where 
\begin{equation}
\label{T(ijk)}
T_{(ijk)}=\frac{1}{6}\left(T_{ijk}+T_{jik}+T_{ikj}+T_{jki}+T_{kij}+T_{kji}\right).
\end{equation}
is a totally symmetric function of quark masses. The sum has $3+3\times 6+6=27$ terms in all, as it should be. Now we make use Eq.\,(\ref{D15}) to obtain 
\begin{equation}
\sum_{i\neq j} J_{iij}T_{(iij)}=\sum_{i<j} \left(J_{ij} T_{iij}^+ +R_{ij} T_{iij}^- \right), 
\end{equation}
where $T_{iij}^\pm =T_{(iij)}\pm T_{(jji)}$.
As a result we finally have 
\begin{eqnarray}
\label{3ind}
\sum_{i,j,k} J_{ijk} \, T_{ijk} &=& \sum_i J_{i}T_{iii}+3\sum_{i<j} \left( J_{ij} T_{iij}^+ +R_{ij} T_{iij}^- \right) \nonumber \\ 
&+& 6 J_{123} T_{(123)}.
\end{eqnarray}

The next sum describes the contribution of the fermion box diagrams with external spin-1 fields to the coefficient $b_2(x,x)$. Here we have 
\begin{eqnarray}
\label{4in}
&&\sum_{i,j,k,l} J_{ijkl} T_{ijkl}=\sum_{i,j,k,l} J_{ijkl} T_{(ijkl)} \nonumber \\
&&=\sum_i J_{i}T_{iiii}+4 \sum_{i\neq j} J_{iiij} T_{(iiij)} +6 \sum_{i< j} J_{iijj} T_{(iijj)} \nonumber \\
&&+12 \sum_{i\neq j\neq k \atop j<k} J_{iijk} T_{(iijk)}.
\end{eqnarray}   
This formula is written for the specific case when each of the indices $i,j,k,l$ takes only three possible values $1,2$, or $3$ (we consider fermion fields with three flavors), so $T_{ijkl}$ has at least two equal indices. As a result, the sum (\ref{4in}) has $3+4\times 6 +6\times 3+12\times 3=81$ terms, and we need to know only the following coefficients 
\begin{eqnarray}
\label{symt4}
T_{(iiij)}&=&\frac{1}{4} \left(T_{iiij}+T_{iiji}+T_{ijii}+T_{jiii}\right), \\
T_{(iijj)}&=&\frac{1}{6} \left(T_{iijj}+T_{jjii}+T_{ijij}+T_{jiji} +T_{ijji}+T_{jiij} \right), \nonumber \\
T_{(iijk)}&=&\frac{1}{12} \left(T_{iijk}+T_{iikj}+T_{kiij}+T_{jiik}+T_{jkii}+T_{kjii} \right. \nonumber \\
&+& \left. T_{ijik}+T_{ikij}+T_{jiki}+T_{kiji}+T_{ijki}+T_{ikji} \right). \nonumber
\end{eqnarray}  

With the use of (\ref{iiij-}) and (\ref{iiij+}) one can set apart the divergent part of the proper-time integral $J_{iiij}$ in the form $J_{ij}$, and as a result to obtain that  
\begin{eqnarray}
\sum_{i\neq j} J_{iiij} T_{(iiij)} &=& \sum_{i<j} \left(J_{ij}+\frac{1}{2} S_{ij} \right)T^+_{iiij} \nonumber \\
&+& \frac{3}{2} \sum_{i<j} R_{ij} T^-_{iiij},
\end{eqnarray}
where $T_{iiij}^\pm=T_{(iiij)}\pm T_{(jjji)}$, and $S_{ij} = S(m_i^2,m^2_j)$.

Thus, finally (\ref{4in}) can be expressed as a sum
\begin{eqnarray}
\label{4inf}
&&\sum_{i,j,k,l} J_{ijkl} T_{ijkl} =\sum_i J_{i}T_{iiii}\nonumber \\
&&+4\sum_{i<j}  \left[ 
\left(J_{ij}+\frac{1}{2}S_{ij}  \right)T^+_{iiij}
+\frac{3}{2}  R_{ij} T^-_{iiij} \right] \nonumber \\
&&+6   \sum_{i< j} \left(J_{ij}- S_{ij} \right) T_{(iijj)}  
  +12\!\!\! \sum_{i\neq j\neq k, \atop j<k}\!\!\! J_{iijk} T_{(iijk)}.
\end{eqnarray}   

\section{Properties of t-coefficients}
\label{app6}
It is the purpose of this Appendix to establish some important properties of t-coefficients which have been used in the computation of the effective action. 

We begin from the coefficient $\hat t_{ijk}$ given by Eq.\,(\ref{coeff-hatt}). For that we need the following expressions
\begin{eqnarray}
&&\!\!\!\!\!\!\!\!\!\!\!\!\! d^\alpha_{ij}=\delta_{ij} \partial^\alpha +i\Gamma^\alpha_{ij}, \\
&&\!\!\!\!\!\!\!\!\!\!\!\!\!\! (d^2)_{ij}=\sum_k d^\alpha_{ik}d^\alpha_{kj} \nonumber \\
&&\ =\delta_{ij}\partial^2 +2i\Gamma_{ij}^\alpha\partial^\alpha +i(\partial^\alpha\Gamma^\alpha_{ij}) -(\Gamma^2)_{ij}.
\end{eqnarray} 
Then for the products we have 
\begin{eqnarray}
&& d^\alpha_{ij}d^\alpha_{jk}(d^2)_{ki} = \delta_{jk} (\partial\Gamma_{ij})\left[ (\partial\Gamma_{ki}) +i (\Gamma^2)_{ki}\right]  \nonumber \\
&&\ \ \ \ \ \ \ \ \ \ \ \ \ \ \ \  -\Gamma^\alpha_{ij} \Gamma^\alpha_{jk}\left[i(\partial \Gamma_{ki})-(\Gamma^2)_{ki}\right],  \\
&& d^\alpha_{jk}(d^2)_{ki}d^\alpha_{ij} = \delta_{ki} (\partial^\beta\Gamma^\alpha_{jk})(\partial^\beta\Gamma^\alpha_{ij}) \nonumber \\
&&\ \ \ \ \ \ \ \ \ \ \ \ +i\Gamma^\beta_{ki}(\Gamma^\alpha_{ij}\! \stackrel{\leftrightarrow}{\partial^\beta}\! \Gamma^\alpha_{jk} ) 
+\Gamma^\alpha_{ij}\Gamma^\alpha_{jk}(\Gamma^2)_{ki}, \\
&& (d^2)_{ki}d^\alpha_{ij}d^\alpha_{jk} = \delta_{ij} (\partial\Gamma_{jk})\left[ (\partial\Gamma_{ki}) -i (\Gamma^2)_{ki}\right]  \nonumber \\
&&\ \ \ \ \ \ \ \ \ \ \ \ \ \ \ \  +\Gamma^\alpha_{ij} \Gamma^\alpha_{jk}\left[  i(\partial \Gamma_{ki})+(\Gamma^2)_{ki}\right],  
\end{eqnarray}
where a total divergence and open derivatives on the right hand side were omitted. It gives
\begin{eqnarray}
\label{hattijk}
&&\hat t_{ijk}=\frac{1}{3}\left[ \frac{}{} \delta_{ij} (\partial\Gamma_{ki}) (\partial\Gamma_{jk}) 
                    +  \delta_{jk} (\partial\Gamma_{ij}) (\partial\Gamma_{ki}) \right.  \nonumber \\
&&+  \delta_{ki} (\partial^\beta\Gamma^\alpha_{jk}) (\partial^\beta\Gamma^\alpha_{ij}) 
    +i(\delta_{jk} \partial\Gamma_{ij}-\delta_{ij} \partial\Gamma_{jk})(\Gamma^2)_{ki} \nonumber \\
&& \left. +i\Gamma^\beta_{ki} (\Gamma^\alpha_{ij}\! \stackrel{\leftrightarrow}{\partial^\beta}\! \Gamma^\alpha_{jk} ) \right]
+\Gamma^\alpha_{ij} \Gamma^\alpha_{jk} (\Gamma^2)_{ki}\,.
\end{eqnarray}

From this formula, in particular, one finds 
\begin{eqnarray}
\label{hatt+iij}
&&3\hat t^+_{iij}=\hat t_{iij} +\hat t_{iji}+\hat t_{jii} +   (i\leftrightarrow j) = \frac{2}{3} (\partial\Gamma_{ij})(\partial\Gamma_{ji}) \nonumber \\
&&+\frac{1}{3}(\partial^\beta\Gamma^\alpha_{ij})(\partial^\beta\Gamma^\alpha_{ji}) 
+ \frac{i}{3}\left[F^{\alpha\beta}_{ij} \Gamma^\alpha_{ji} (\Gamma^\beta_{ii} - \Gamma^\beta_{jj} )\right. \nonumber \\
&&\left. +F^{\alpha\beta}_{ii}\Gamma^\alpha_{ij} \Gamma^\beta_{ji}\right] + \Gamma^\alpha_{ji} (\Gamma^\alpha_{ii} +\Gamma^\alpha_{jj})(\Gamma^2)_{ij}  +\Gamma^\alpha_{ij} \Gamma^\alpha_{ji}(\Gamma^2)_{ii}\nonumber \\
&&  +(i\leftrightarrow j).
\end{eqnarray}
 and 
\begin{eqnarray}
\label{coeff-t-iij}
&&3\hat t^-_{iij}=\hat t_{iij} +\hat t_{iji}+\hat t_{jii} -(i\leftrightarrow j) = \frac{2i}{3} (\partial\Gamma_{ji})(\Gamma^2)_{ij} \nonumber \\
&&+ \frac{i}{3}\left[ F^{\alpha\beta}_{ij} \Gamma^\alpha_{ji} ( \Gamma^\beta_{ii} + \Gamma^\beta_{jj} )
+F^{\alpha\beta}_{ii}\Gamma^\alpha_{ij} \Gamma^\beta_{ji}\right] \nonumber \\
&& + \Gamma^\alpha_{ji} (\Gamma^\alpha_{ii} -\Gamma^\alpha_{jj})(\Gamma^2)_{ij}+\Gamma^\alpha_{ij} \Gamma^\alpha_{ji}(\Gamma^2)_{ii} -(i\leftrightarrow j).
\end{eqnarray}

From (\ref{hattijk}) it follows also that in the case of three different masses, $m_1\neq m_2\neq m_3$, the t-coefficient $\hat t_{(123)}$ is 
\begin{equation}
\label{hatttotsym}
\hat t_{(123)}=\frac{1}{6}\sum_{i\neq j\neq k} \left[ \frac{i}{3}\Gamma^\alpha_{ij} \Gamma^\beta_{jk} F^{\alpha\beta}_{ki}   +
\Gamma^\alpha_{ij} \Gamma^\alpha_{jk} (\Gamma^2)_{ki}\right].
\end{equation}

Now consider the coefficient $t_{ijkl}$ given by Eq.(\ref{coeff-hatt}). Up to a total divergence it can be rewritten as
\begin{eqnarray}
\label{t_ijkl}
&&t_{ijkl}=\frac{1}{4}\delta_{\alpha\beta\gamma\sigma} \left[4\Gamma^\alpha_{ij}\Gamma^\beta_{jk}\Gamma^\gamma_{kl}\Gamma^\sigma_{li}   \right. \nonumber \\
&&+\delta_{jk}\delta_{kl} (\partial^\beta \Gamma^\alpha_{ij})      (\partial^\gamma \Gamma^\sigma_{li}) 
     +\delta_{kl}\delta_{li} (\partial^\gamma \Gamma^\beta_{jk})   (\partial^\sigma \Gamma^\alpha_{ij})  \nonumber \\
&&+\delta_{li}\delta_{ij}  (\partial^\sigma \Gamma^\gamma_{kl}) (\partial^\alpha \Gamma^\beta_{jk})
    +\delta_{ij}\delta_{jk} (\partial^\alpha \Gamma^\sigma_{li})      (\partial^\beta \Gamma^\gamma_{kl})  \nonumber \\      
&&+i\delta_{kl}\Gamma^\alpha_{ij} (\Gamma^\sigma_{li}\! \stackrel{\leftrightarrow}{\partial^\gamma}\!  \Gamma^\beta_{jk})  +i\delta_{jk}\Gamma^\sigma_{li} (\Gamma^\gamma_{kl}\! \stackrel{\leftrightarrow}{\partial^\beta}\! \Gamma^\alpha_{ij}) \nonumber \\
&& +\left. i\delta_{li}\Gamma^\beta_{jk} (\Gamma^\alpha_{ij}\! \stackrel{\leftrightarrow}{\partial^\sigma}\!  \Gamma^\gamma_{kl}) +i\delta_{ij}\Gamma^\gamma_{kl} (\Gamma^\beta_{jk}\! \stackrel{\leftrightarrow}{\partial^\alpha}\! \Gamma^\sigma_{li}) \right].
\end{eqnarray}
Hence we have 
\begin{equation}
t_{iiii}=2(\partial\Gamma_{ii})^2 +(\partial^\alpha\Gamma^\beta_{ii})^2 +3(\Gamma^\alpha_{ii}\Gamma^\alpha_{ii})^2. 
\end{equation}

Further, from the definition of the coefficients and the property of the cyclic permutation, it follows that
\begin{eqnarray}
\label{t3-1}
&& t_{iiij}=t_{iiji}=t_{ijii}=t_{jiii}=t_{(iiij)}\,, \\
\label{t2-2}
&& t_{iijj}=t_{ijji}=t_{jjii}=t_{jiij}\,, \\
\label{t1-1-1-1}
&& t_{ijij}=t_{jiji}\,.
\end{eqnarray}
Therefore
\begin{eqnarray}
\label{t_iiij}
&& t_{(iiij)}=\frac{1}{4}\delta_{\alpha\beta\gamma\sigma}\left[(\partial^\alpha\Gamma^\sigma_{ji})(\partial^\beta\Gamma^\gamma_{ij})+ 4\Gamma^\alpha_{ii}\Gamma^\beta_{ii}\Gamma^\gamma_{ij}\Gamma^\sigma_{ji}     \right. \nonumber \\
&& +\left. i(\partial^\alpha\Gamma^\sigma_{ji})\Gamma^\beta_{ii}\Gamma^\gamma_{ij}-i (\partial^\beta\Gamma^\gamma_{ij})\Gamma^\sigma_{ji}\Gamma^\alpha_{ii}  \right] \nonumber \\
&&=\frac{1}{4} \left\{\frac{}{} 2(\partial\Gamma_{ij})(\partial\Gamma_{ji}) +(\partial^\alpha\Gamma^\beta_{ij})(\partial^\alpha\Gamma^\beta_{ji})   \right.  \nonumber \\
&&+\left. i \Gamma^\beta_{ii}\left[ (\Gamma^\beta_{ij}\! \stackrel{\leftrightarrow}{\partial^\alpha}\! \Gamma^\alpha_{ji} )
 -(\Gamma^\beta_{ji}\! \stackrel{\leftrightarrow}{\partial^\alpha}\! \Gamma^\alpha_{ij} )
+(\Gamma^\alpha_{ij}\! \stackrel{\leftrightarrow}{\partial^\beta}\! \Gamma^\alpha_{ji} )\right] \right\}  \nonumber \\
&&+(\Gamma_{ii}\Gamma_{ii})(\Gamma_{ij}\Gamma_{ji})+2(\Gamma_{ii}\Gamma_{ij})(\Gamma_{ii}\Gamma_{ji}).
\end{eqnarray}
From (\ref{t_iiij}), we obtain 
\begin{eqnarray}
\label{coeff-t-pm-iiij}
&&t^\pm_{iiij}=\delta_{\pm +}\left[(\partial\Gamma_{ij})(\partial\Gamma_{ji}) +\frac{1}{2} (\partial^\alpha\Gamma^\beta_{ij})(\partial^\alpha\Gamma^\beta_{ji}) \right] \nonumber \\
&&  +\frac{i}{4}\!\left(\Gamma^\beta_{ii} \mp \Gamma^\beta_{jj}\right)\!\! \left[(\Gamma^\beta_{ij}\! \stackrel{\leftrightarrow}{\partial^\alpha}\! \Gamma^\alpha_{ji} )
              -(\Gamma^\beta_{ji}\! \stackrel{\leftrightarrow}{\partial^\alpha}\! \Gamma^\alpha_{ij} )
             +(\Gamma^\alpha_{ij}\! \stackrel{\leftrightarrow}{\partial^\beta}\! \Gamma^\alpha_{ji} ) \right] \nonumber \\
&&+\left[(\Gamma_{ii}\Gamma_{ii})(\Gamma_{ij}\Gamma_{ji})+2(\Gamma_{ii}\Gamma_{ij})(\Gamma_{ii}\Gamma_{ji}) \pm (i\leftrightarrow j)\right],    
\end{eqnarray}
where $\delta_{++}=1$, and $\delta_{-+}=0$.

Next, we calculate the coefficient $t_{(iijj)}$. Using the formulas (\ref{symt4}), (\ref{t2-2}) and (\ref{t1-1-1-1}) it is easy to establish that 
\begin{equation}
t_{(iijj)}=\frac{1}{3}\left(2t_{iijj}+t_{ijij}\right).
\end{equation}
The formula (\ref{t_ijkl}) gives now 
\begin{eqnarray}
\label{coeff-t-(iijj)}
&&t_{(iijj)}= \frac{1}{3}\left\{ \frac{}{}\!\! (\Gamma_{ii}\Gamma_{ij})(\Gamma_{jj}\Gamma_{ji}) + (\Gamma_{ij}\Gamma_{jj})(\Gamma_{ji}\Gamma_{ii})\right.\nonumber \\
&& +(\Gamma_{ii}\Gamma_{jj})(\Gamma_{ij}\Gamma_{ji})+(\Gamma_{ij}\Gamma_{ji})^2+ \frac{1}{2} (\Gamma_{ij}\Gamma_{ij})(\Gamma_{ji}\Gamma_{ji}) \nonumber \\
&&+\frac{i}{2}\Gamma^\alpha_{ii}\left[ (\Gamma^\alpha_{ji}\! \stackrel{\leftrightarrow}{\partial^\beta}\! \Gamma^\beta_{ij})+(\Gamma^\beta_{ji}\! \stackrel{\leftrightarrow}{\partial^\alpha}\! \Gamma^\beta_{ij} ) + (\Gamma^\beta_{ji}\! \stackrel{\leftrightarrow}{\partial^\beta}\! \Gamma^\alpha_{ij} ) \right] \nonumber \\
&&+\left. (i\leftrightarrow j) \frac{}{}\!\! \right\}.
\end{eqnarray}

Finally, consider the coefficient $t_{(iijk)}$. In accord with the definition (\ref{symt4}) and the cyclic property (\ref{coeff-hatt}) we have 
\begin{eqnarray}
\label{t_(iijk)}
&&t_{(iijk)}=\frac{1}{3} \left(t_{iijk} +t_{iikj}+t_{ikij}\right) \nonumber \\
&&=\frac{1}{3}\delta_{\alpha\beta\gamma\sigma} \left[ \Gamma^\alpha_{ii} \Gamma^\beta_{ij} \Gamma^\gamma_{jk} \Gamma^\sigma_{ki}  + \Gamma^\alpha_{ii} \Gamma^\beta_{ik} \Gamma^\gamma_{kj} \Gamma^\sigma_{ji} 
+\Gamma^\alpha_{ij} \Gamma^\beta_{ji} \Gamma^\gamma_{ik} \Gamma^\sigma_{ki}\right. \nonumber \\
&&\left. +\frac{i}{4} \Gamma^\gamma_{jk} \left(\Gamma^\beta_{ij} \! \stackrel{\leftrightarrow}{\partial^\alpha}\! \Gamma^\sigma_{ki}\right) + \frac{i}{4} \Gamma^\gamma_{kj} \left(\Gamma^\beta_{ik} \! \stackrel{\leftrightarrow}{\partial^\alpha}\! \Gamma^\sigma_{ji}\right)\right].
\end{eqnarray}

In conclusion, we recall that the coefficients depending on $Y$: $t_{ij}$ and $t_{ijk}$ were calculated in the main part of the work.


\end{document}